%% file: PAPER_PTRSA_Geo_arxiv_r1.tex
\documentclass[3p]{scrartcl}
\usepackage[hmarginratio=2:1,top=25mm,left=20mm,columnsep=1cm]{geometry} %

\input{defs_arxiv} 
\setcapindent{0pt}

\begin{document}

\title{\bfseries A unified first order hyperbolic model for nonlinear dynamic rupture processes in 
diffuse 
fracture zones}
\author{
A.-A. Gabriel\thanks{Ludwig-Maximilians-Universit\"at M\"unchen, Theresienstr. 41, 80333 M\"unchen, 
Germany, \href{mailto:gabriel@geophysik.uni-muenchen.de}{email}}, 
\quad
D. Li\samethanks[1], 
\quad
S. Chiocchetti\thanks{Laboratory of Applied Mathematics, University of Trento,
	Via Mesiano 77, 38123, Trento, Italy}, 
\quad 
 M. Tavelli\thanks{Free University of Bolzano, Universit\"atplatz 1 , 39100 Bolzano, Italy
 }$ ^{\ \ ,} $\samethanks[2],
\\[2mm]
 I. Peshkov\samethanks[2], 
 \quad
 E. 
Romenski\thanks{Sobolev Institute of Mathematics, 4 Acad. Koptyug Avenue, 630090,
	Novosibirsk, Russia,}$ ^{\ \ ,} $\samethanks[2], 
\quad
M. Dumbser\samethanks[2]}
\thanksmarkseries{arabic}




\date{}
\maketitle

\begin{abstract}
	\noindent
Earthquake fault zones are more complex, both geometrically and rheologically, than an idealised infinitely thin plane embedded in linear elastic material. To incorporate nonlinear material behaviour, natural complexities and multi-physics coupling within and outside of fault zones, here we present a first order hyperbolic and thermodynamically compatible mathematical model for a continuum in a gravitational field which provides a unified description of nonlinear elasto-plasticity, material damage and of viscous Newtonian flows with phase transition between solid and liquid phases. The fault geometry and secondary cracks are described via a scalar function $\xi \in [0,1]$ that indicates the local level of material damage. The model also permits the representation of arbitrarily complex geometries via a diffuse interface approach based on the solid volume fraction function $\alpha \in [0,1]$. 
Neither of the two scalar fields $\xi$ and $\alpha$ needs to be mesh-aligned, allowing thus faults and cracks with complex topology and the use of adaptive Cartesian meshes (AMR). 
The model shares common features with phase-field approaches, but substantially extends them. We show a wide range of numerical applications that are relevant for dynamic earthquake rupture in fault zones, including the co-seismic generation of secondary off-fault shear cracks, tensile rock fracture in the Brazilian disc test, as well as a natural convection problem in molten rock-like material.  
\end{abstract}


%

\blfootnote{\vphantom{\Huge ${A^{A}}^{A}$}Accepted to \textit{Phil. Trans. R. 
Soc. A}, \url{doi.org/10.1098/rsta.2020.0130}}

\section{Introduction}
Multiple scales, multi-physics interactions and nonlinearities govern earthquake source processes, 
rendering the  
understanding of how faults slip a grand challenge of seismology \cite{Forsyth2009, Nielsen2017}.
Over the last decades, earthquake rupture dynamics have been commonly modeled as a sudden 
displacement 
discontinuity across a simplified 
(potentially heterogeneous) surface of zero thickness in the framework of elastodynamics 
\cite{Andrews1976}. 
Such earthquake models are commonly forced to distinguish artificially between on-fault frictional 
failure and the off-fault response of rock.
Here, we model natural fault damage zones \cite{Chester1993, Mitchell2009} by adopting a diffuse 
crack representation. 

In recent years, the core assumption that faults behave as infinitely thin planes has been 
challenged \cite{Wibberley2008}.
Efforts collapsing the dynamics of earthquakes to single interfaces may miss important physical 
aspects governing fault-system behaviour such as
complex volumetric failure patterns observed in recent well-recorded large and small 
earthquakes \cite{Ross2019,Cheng2018} 
as well as in laboratory experiments \cite{Passelegue2016}.
However the mechanics of fault and rupture dynamics in \emph{generalized nonlinear 
visco-elasto-plastic materials} are challenging to incorporate in earthquake modeling.
{Earthquakes propagate as frictional shear fracture of brittle solids under compression 
along pre-existing weak interfaces (fault zones),
	a problem which is mostly unsolvable analytically.
	For numerical modeling, dynamic earthquake rupture is often treated as a nonlinear boundary 
	condition\footnote{
		Faults are then idealised as two matching surfaces in unilateral contact not allowed to 
		open or interpenetrate and typically implemented by splitting the fault interface 
		\cite{Day2005}.} in terms of contact and friction, 
	coupled to seismic wave propagation in linear elastic material.
	The evolving displacement discontinuity across the fault is defined as the earthquake induced 
	slip.
	Typically, the material surrounding the fault is assumed to be linear, isotropic and elastic, 
	with all nonlinear complexity collapsed into the boundary condition definition of fault 
	friction \cite[e.g.]{Gabriel2012}, which take the form of empirical laws describing shear 
	traction bounded by the fault strength. 
	In an elastic framework, high stress concentrations develop at the rupture front.  
	{The corresponding inelastic off-fault energy dissipation (\textit{off-fault damage}) 
		and its feedback on rupture propagation \cite{Kikuchi1975} can be modelled in form of 
		(visco-)plasticity of Mohr-Coulomb or Drucker-Prager type \cite{Andrews2005, 
		Templeton2008}, 
		a continuum damage rheology which may account for high strain rate effects 
		\cite{Lyakhovsky2005, Bhat2012, Thomas2018},
		or explicit secondary tensile and shear fracturing \cite{Yamashita2000, Dalguer2003, 
		Okubo2019}.}}

Numerical modeling of crack propagation has been a long-standing problem not only in seismology 
but also in computational mechanics. {Emerging approaches in modeling fracture and rupture 
dynamics include 
	phase-field and varifold-based representations of cracks to tackle the} 
major difficulty of the introduction of strong discontinuities in the displacement field in the 
vicinity of the crack. Current state-of-the-art methods in earthquake rupture dynamics 
\cite{Harris2018} require explicit fracture aligned meshing,  
thus, generally ({with recent exceptions \cite{Okubo2020}}) require fractures to be 
predefined, and typically only permit small deformations. 
Using highly efficient software implementations of this approach large-scale earthquake modeling 
is possible\cite{Cui2013,Heinecke2014,Uphoff2017}.
Alternative spatial discretizations which allow representing strong discontinuities 
at the sub-element level, such as the eXtended Finite Element Method (XFEM) 
\cite{Moes1999}, introduce singularities when an interface intersects a cell, but 
are quite difficult to implement in an efficient manner.

In distinction, \emph{diffuse interface approaches} ``smear out'' sharp cracks 
via a smooth but rapid transition between intact and 
fully damaged material states \cite{Bourdin2000,deBorst2004,Mirzabozorg2005}. 
Within various diffuse interface approaches, the most 
popular one is the \emph{phase field approach}, which 
allows to model complicated fracture processes, including spontaneous crack initiation, 
propagation, merging, 
and branching, in general situations and for 3D geometries.
Critical ingredients of the phase-field formulation are rooted in fracture mechanics, 
specifically by incorporating a critical fracture energy (from Griffith's theory 
\cite{Griffith1921}), 
which is translated into the regularized continuum gradient damage mechanics 
\cite{Miehe2015}. 
Several theoretical methods have been 
recently proposed for shear fracture (\cite[e.g.]{Karma2011} for mode III) which is dominating 
earthquake processes.
Phase-field models have also been successfully applied for brittle fracture in rock-like materials 
\cite{Zhang2017} on small scales (mm's of slip). 

%
The material failure model discussed in this paper also belongs to the class of diffuse interface 
models in which the damaged material or a crack is considered as 
another ``\emph{phase}'' of the material and represented by a continuous scalar field $ \xi\in[0,1] 
$, called 
the \emph{damage variable}. As in phase field approaches, a 
crack or failure front is 
represented not as a discontinuity of zero thickness but as a \emph{diffuse interface} across which 
$ \xi $ changes continuously from 0 (intact material) to 1 (fully damaged material) resulting in 
gradual but rapid degradation of material stiffness. Despite this conceptual similarity, the 
model developed here is very different from the phase field models. 
An important feature of the phase field models is the presence of the non-local regularization term 
$ 
\sim \Vert\nabla \phi \Vert^2 $ {in the free energy, with $ \phi $ being the phase field}. 
Without 
such a 
regularization 
term, the numerical treatment of a phase field model is problematic due to numerical 
instabilities and mesh dependency of the numerical solution. 
This indicates the 
ill-posedness 
of the underlying governing PDEs, {e.g. see \cite{Lorentz2005,Steinmann2001}}.
In contrast, the model developed here originating from \cite{Resnyansky2003,Romenskii2007} does not 
require non-local 
regularization terms\footnote{Non-local 
	terms can be introduced in our theory if it is physically motivated, e.g. see 
	\cite{PRD-Torsion2019,Chiocchetti2020}} and is formulated 
based on the thermodynamically compatible continuum mixture theory 
\cite{Romenski2007,RomDrikToro2010} which 
results in a \emph{first-order symmetric hyperbolic} 
governing 
PDE system and thus is intrinsically well-posed, at least locally in time. Mathematical 
regularity of the model is supported by the stability of the hereafter  
presented numerical results, including a mesh convergence analysis (see Sec.\,\ref{sec:results}). 
More generally, the 
developed model belongs to the class of Symmetric Hyperbolic and Thermodynamically Compatible 
(SHTC) 
equations \cite{God1961,GodRom1996a,Rom1998,SHTC-GENERIC-CMAT}.
Apart from the PDE type used (the phase field models are formulated as second-order Allen-Cahn-type 
\cite{Ambati2014,Gomez2017a} or fourth-order Cahn-Hilliard-type 
\cite{Eastgate2002,Borden2014,Chiarelli2017} parabolic PDEs), there is also an important conceptual 
difference between the developed 
mixture type approach and the phase field approaches. In the latter, the phase 
transformation is entirely controlled by the free energy functional, which usually consists of 
three terms:
$
\Psi(\boldsymbol{\varepsilon},\phi,\nabla\phi) = \Psi_1(\boldsymbol{\varepsilon},\phi) + 
\Psi_2(\phi) + \Psi_3(\nabla\phi),
$ 
where $ \boldsymbol{\varepsilon} $ is the small elastic strain tensor, $ \Psi_1 $ is the elastic 
energy which comprises a degradation function, $ \Psi_2 $ is the damage potential (usually a 
double-well potential but also single-well potentials are used \cite{Levitas2018}), and $ \Psi_3 $ 
is the regularization term. In our approach, only an energy equivalent to $ 
\Psi_1(\boldsymbol{\varepsilon},\phi) $ is used \cite{Romenskii2007,GPRCrack}, while the 
phase-transition is described in the context of irreversible thermodynamics and is controlled by a 
dissipation potential which is usually a highly nonlinear function of state variables\footnote{For 
	example, relaxation times may change over several orders of magnitude across the diffuse 
	interface zone.}
\cite{SHTC-GENERIC-CMAT,PKG-Book2018}. Yet, it is
important to emphasize that the irreversible terms controlling the damage are \emph{algebraic} 
source 
terms (no space derivatives), which do not affect the differential operator of the model. This 
greatly simplifies the discretization of the differential terms in the governing PDE, but 
nevertheless requires an 
accurate and robust stiff ordinary differential equation solver \cite{Chiocchetti2020a,GPRCrack} 
for the source terms. 
{Since} the governing PDE system of our theory 
contains only first-order derivatives in space and time, it is possible to use explicit 
time-stepping in the numerical integration \cite{GPRCrack}. In contrast, the second and 
fourth-order phase field PDEs require the use of an implicit time discretization \cite{Gomez2017a}, 
which is more difficult to implement and may not have advantage over explicit methods if the time 
step is 
dictated by the physical time scales, such as in strongly time dependent processes, e.g. fracture 
dynamics and wave propagation. {We note that a hyperbolic reformulation of 
	phase-field models is possible as recently proposed in \cite{Kamensky2018}.}

{
	Alternatively, variational views on fracture mechanics can describe crack nucleation 
	intrinsically
	without a priori failure criteria \cite{Francfort1998, Mariano2010}.
	Accounting for microscopic surface irregularities or line defects can be achieved by combining 
	a sharp interface approach 
	with advanced tools of differential geometry such as \emph{curvature varifolds} 
	\cite{Giaquinta2010}. 
	These ideas can be seen as a natural extension of the pioneering Griffith's theory 
	\cite{Griffith1921} with cracks being 
	represented by almost everywhere differentiable surfaces and evolving Griffith's energies to 
	account for curvature effects.
	In this context, we remark that the model presented here by no means is a complete fracture 
	model. 
	In specific situations requiring a very accurate prediction of the fracture process the merely 
	constitutive 
	capabilities of the present model may not be sufficient. Instead, accounting explicitly for the 
	energy accumulating 
	at the irregularities of the crack surface (e.g., at corners and cusps) or the dynamics of 
	microscopic defects 
	near the crack tip might be required. 
	In the here presented first-order hyperbolic diffuse interface framework, this can be achieved 
	by 
	taking into account higher gradients of the state variables such as curvature and torsion 
	in the form of independent state variables \cite{PRD-Torsion2019,Chiocchetti2020}. 
}


\section{Mathematical model} 
\label{sec:model} 

The continuum model for damage of solids employed in this paper consists of two main ingredients. 
The first ingredient is the damage model proposed by Resnyansky, Romenski and co-authors
\cite{Resnyansky2003,Romenskii2007} which is a continuous damage model with a chemical 
kinetics-type mechanism controlling the damage field $ \xi \in [0,1]$ ($ \xi=0 $ corresponds to the 
intact and $ \xi=1 $ to the fully damaged state), which is interpreted as the concentration of the 
damaged phase. Being a relaxation-type approach, it provides a 
rather universal framework 
for modeling brittle and ductile fracture from a unified non-equilibrium thermodynamics viewpoint, 
according to which these two 
types of fractures can be described by the same constitutive relations (relaxation functions), but 
have different characteristic time scales, e.g. see \cite{GPRCrack}.
The second ingredient is the Eulerian finite strain elastoplasticity model developed by Godunov and 
Romenski in the 1970s \cite{GodRom1972,God1978,Romenski1979}. It was recently realized by Peshkov 
and 
Romenski \cite{PeshRom2014} that the same equations can also be applied to modeling viscous fluid 
flow, as demonstrated by Dumbser \textit{et al} in \cite{DumbserPeshkov2016} and thus, 
this model represents a unified formulation of continuum fluid and solid mechanics. In the 
following we shall refer to it as Godunov--Peshkov--Romenski (GPR) model. 
Being essentially an inelasticity theory, the GPR model provides a unified framework for continuous 
modeling of 
potentially arbitrary rheological responses of materials, and in particular of inelastic properties 
of the damaged material. This, in turn, can be used for modeling of complex frictional rheology in 
fault zones in geomaterials, see Sec.\,\ref{sec:results}. 
For further details on the GPR model, the reader is referred to 
\cite{PeshRom2014,DumbserPeshkov2016,Jackson2019a,FrontierADERGPR,Rom1998}.



%
Our diffuse interface formulation for moving nonlinear elasto-plastic solids of arbitrary geometry 
and at large strain is given by the following PDE system in Eulerian coordinates: 
\begin{subequations}\label{eqn.GPR}
	\begin{align}
		& \pd_t \alpha +v_k \pd_k \alpha =0, \qquad  
		\pd_t \bar{\rho} + \pd_k (\bar{\rho} v_k ) =0,
		\label{eqn.alpha.mass} \\[2mm]
		&\pd_t (\bar{\rho} v_i)  + \pd_k \left(  \bar{\rho} v_i v_k + \alpha p 
		\delta_{ik} - \alpha \sigma_{ik}  \right)   = \bar{\rho} g_i, 
		\label{eqn.momentum} \\[2mm]
		& \pd_t A_{ik} + \pd_k (A_{im}v_m) + 
		v_m \left( \pd_m A_{ik} - \pd_k A_{im}  \right) = 
		-\theta_1^{-1}(\tau_1) E_{A_{ik}}, 
		\label{eqn.A} \\[2mm]
		& \pd_t J_k + \pd_k \left( v_m J_m + T \right) + 
		v_m \left( \pd_m J_{k} - \pd_k J_{m} \right) = - 
		\theta_2^{-1}(\tau_2) E_{J_k},  
		\label{eqn.J} \\[2mm]
		& \pd_t \xi + v_k \pd_k \xi = - \theta E_\xi, 
		\label{eqn.xi} \\[2mm]
		& \pd_t (\bar{\rho} S) + \pd_k \left( \bar{\rho} S v_k + \bar{\rho} E_{J_k} \right) = 
		\bar{\rho} \, (\alpha T)^{-1} \left( \theta_1^{-1} E_{A_{ik}} E_{A_{ik}} + \theta_2^{-1} 
		E_{J_k} 
		E_{J_k}  + 
		\theta E_\xi E_\xi \right) \geq 0,  
		\label{eqn.entropy} \\[2mm]
		& \pd_t (\bar{\rho} {E})  + \pd_k \left( v_k \bar{\rho} E +v_i(\alpha p\delta_{ik} - \alpha 
		\sigma_{ik}) 
		\right) = \bar{\rho} g_i v_i,  
		\label{eqn.energy} 
	\end{align}
\end{subequations} 
where we use the Einstein summation convention over repeated indices 
and $ \pd_t = \pd/\pd t $, $ \pd_k = \pd/\pd x_k $. Here, \eqref{eqn.alpha.mass}$ _1 $ is the 
evolution equation for the colour function $\alpha$ that 
is needed in the diffuse interface approach (DIM) as introduced in 
\cite{Tavelli2019,FrontierADERGPR} for the description of solids of arbitrary geometry ($ \alpha = 
1 $ inside of the solid body and $ \alpha = 0  $ outside); 
{$ \bar{\rho} =\alpha \rho$} and \eqref{eqn.alpha.mass}$ _2 $ is the mass conservation law 
with 
$\rho$ being the material density; 
\eqref{eqn.momentum} is the momentum conservation law and $v_i$ is the velocity field; 
\eqref{eqn.A} is the evolution equation for the \textit{distortion field} $\AA = [A_{ik}]$, which 
is 
the main field in the GPR model and can be viewed as the field of \emph{local basis 
	triads}\footnote{Global 
	deformation can not be restored from the local triad since they represent only local 
	deformation and thus, incompatible deformation.} representing the deformation and orientation 
of an 
infinitesimal material element \cite{PRD-Torsion2019,PeshRom2014,DumbserPeshkov2016}; 
\eqref{eqn.J} is the evolution equation for the specific thermal impulse $J_k$, describing the heat 
conduction in the matter via a hyperbolic (non Fourier--type) model; \eqref{eqn.xi} is the 
evolution equation for the material damage variable $\xi\in[0,1]$, where $\xi=0$ indicates fully 
intact material and $\xi=1$ fully damaged material. Finally, \eqref{eqn.entropy} is the entropy 
evolution equation with the positive source product on the right hand-side (second law of 
thermodynamics) and \eqref{eqn.energy} is the energy conservation law (first law of thermodynamics).
Other thermodynamic parameters are defined via the total energy potential 
$E=E(\rho,S,\vv,\AA,\JJ,\xi)$: 
{$p = \rho^2 E_\rho$ is the thermodynamic pressure, 
	$\boldsymbol{\sigma} = [\sigma_{ik}]=[\sigma_{ik}^e+\sigma_{ik}^t]$ is the stress tensor with 
	contributions to the mechanical stress due to tangential $[\sigma_{ik}^e=-\rho A_{ji} 
	E_{A_{jk}}]$ 
	and 
	thermal stress $[\sigma_{ik}^t=\rho J_i E_{J_k}]$ (note that $\boldsymbol{\sigma}^e$ in not 
	necessary 
	trace-free)}, and $T = E_S$ is the temperature. 
{The total mechanical stress tensor is defined as $\boldsymbol{\Sigma} = [{\Sigma}_{ik}] = 
	[-p\delta_{ik} + \sigma_{ik}^e]$, where $\delta_{ik}$ is the Kronecker delta.}
With a state variable in the 
subscript of the energy, we denote partial derivatives, e.g. $ E_\rho = \pd E/\pd\rho $, $ 
E_{A_{ij}} 
= \pd E/\pd A_{ij} $, etc. Furthermore, $g_i$ is the gravitational acceleration vector. Also, 
because 
we are working in an Eulerian frame of reference, we 
need to add transport equations of the type $\pd_t \lambda + v_k\pd_k \lambda = 0 $ to the 
above evolution equations for all the material parameters (e.g., Lam\'e constants) {in case 
	of heterogeneous material properties}, see 
\cite{GPRCrack}.


In order to close the system one must specify the total energy potential as a function 
of the state variables, i.e. $ E = E(\rho,S,\vv,\AA,\JJ,\xi) $. This  potential then 
generates the fluxes (reversible time evolution) and source terms (irreversible time evolution) by 
means of its partial derivatives 
(thermodynamic forces) with respect 
to the state variables.
Here, we make the choice $E=E_1+E_2+E_3$, decomposing the energy into a contribution from the 
microscale $E_1$, the mesoscale $E_2$ and the macroscale $E_3$. The individual contributions read 
as follows:  
\begin{equation}
	E_1 =  K \left( 1 - \rho/\rho_0\right)^2/(2 \rho_0) + c_v T_0\left(\rho/\rho_0 
	\right)\left(e^{S/c_v}-1 \right) + H(T-T_c) h_c, \label{E1a}
\end{equation} 
where $\rho_0$ and $ T_0 $ are the reference mass density and temperature, $h_c$ is the latent 
heat, $T_c$ is the critical temperature at which phase transition occurs, $H(T)$ is the Heaviside 
step function, $ c_v $ is the heat capacity at constant volume. {As a proof of concept, we 
	added the last term in 
	\eqref{E1a} and present a demonstration example of the model's capability to deal with 
	solid-fluid phase transition 
	(melting/solidification) in Section\,5 of the supplementary material. 
	Yet, this corresponds to a simplified (time-independent) modeling of phase transition and 
	will be improved in the future.}
Also, $K(\xi)=\lambda(\xi)+\frac{2}{3}\mu(\xi)$ is the bulk modulus, $\lambda(\xi)$ and $\mu(\xi)$ 
are the two Lam\'e constants that are functions of the damage variable $ \xi $ specified, 
following 
\cite{Resnyansky2003}, as
\begin{equation}
	\lambda(\xi)=K_\i K_\d /\tilde K- 2\mu_\i \mu_\d / (3\tilde{\mu}), \quad 
	\mu(\xi)=\mu_\i \mu_\d / \tilde{\mu},
	\label{Lame}
\end{equation}
where the subscripts $ I $ and $ D $ denote `\emph{intact}' and `\emph{damaged}'
respectively, $K_\i =\lambda_\i +\frac{2}{3}\mu_\i $, $K_\d =\lambda_\d +\frac{2}{3}\mu_\d $, 
$\tilde K=\xi 
K_\i +(1-\xi)K_\d $,
$\tilde \mu=\xi \mu_\i +(1-\xi)\mu_\d $, and it is assumed that the elastic moduli of the intact 
material $\lambda_\i $, 
$\mu_\i $ and of the fully damaged material $\lambda_\d $, $\mu_\d $ are known. 

The macro-scale energy is the specific kinetic energy $
E_3 =  \frac{1}{2} v_i v_i $. Finally, $ E_2 $ reads
\begin{equation} 
	E_2 =  \frac{1}{4} c_s^2 \mathring{G}_{ij} \mathring{G}_{ij} + \frac{1}{2} c_h^2 J_i J_i, 
	\label{E2a}
\end{equation} 
where $c_s(\xi)=\sqrt{\mu(\xi)/\rho_0}$ is the shear sound speed and $c_h$ is related to the speed 
of heat waves 
in the medium (also called the second sound \cite{Peshkov1944}, or the speed of phonons).  
$\mathring{G}_{ik} = G_{ik} - \frac{1}{3} G_{jj} \, \delta_{ik}$ is the deviator of the 
Finger (or metric) tensor $G_{ik} = A_{ji} A_{jk}$ that characterizes the elastic deformation of 
the medium.

The dissipation in the system includes three irreversible processes that raise the entropy:
the strain relaxation (or shear stress relaxation) characterized by the scalar function 
$\theta_1(\tau_1) > 0$ in \eqref{eqn.A} depending on the relaxation time $\tau_1$, the heat flux 
relaxation 
characterized by 
$\theta_2(\tau_2) > 0$ in \eqref{eqn.J}, depending on the relaxation time $\tau_2$, and the 
chemical kinetics like 
process governing the transition from the intact to damaged state and controlled by the function $ 
\theta $ in \eqref{eqn.xi}.



The main idea of the diffuse interface approach to fracture is to consider the material element as 
a \textit{mixture} of the \textit{intact} and the \emph{fully damaged} phases.  
These two phases have their own independent material parameters and closure relations, such 
as functions characterizing the rate of strain relaxation. The strain relaxation approach in the 
framework of the unified hyperbolic continuum mechanics model \cite{PeshRom2014,DumbserPeshkov2016} 
represented by the evolution equation for the distortion field $ \AA $ 
allows to assign potentially arbitrary rheological properties to the damaged and intact states. In 
particular, the intact material can be considered as an elastoplastic solid, while the damaged 
phase 
can be a fluid, e.g. a Newtonian fluid (see Sec.\,\ref{sec:results}\ref{sec.convection}) or 
viscoplastic fluid, 
which can be used for modeling of 
in-fault friction, for example. 
Yet, in this paper, we do not use an individual distortion evolution equation for each phase, 
but we employ the mixture approach \cite{Resnyansky2003,Romenskii2007}, and we use a single 
distortion field representing the local deformation of the mixture element, while the individual 
rheological properties of the phases are taken into account via the dependence of the relaxation 
time $\tau_1$ on the damage variable $\xi$ as follows: 
\begin{equation}\label{tau}
	\tau_1= \left((1-\xi)/\tau_\i  + 
	\xi/\tau_\d\right)^{-1},
\end{equation}
where $\tau_\i$ and $\tau_\d$ are shear stress relaxation times for the intact and fully damaged 
materials, respectively, which are usually highly nonlinear functions of the parameters of state. 
The particular choice of $\tau_\i$ and $\tau_\d$ that is used in this  paper reads
\begin{equation}\label{taus}
	\tau_\i=\tau_{\i0}\exp(\alpha_\i -\beta_\i (1-\xi)Y), \quad \tau_\d=\tau_{\d0}\exp(\alpha_\d 
	-\beta_\d 
	\xi 
	Y),
\end{equation}
where $Y$ is the equivalent stress, while $\tau_{\i0},\alpha_\i 
,\beta_\i $, 
$\tau_{\d0},\alpha_\d ,\beta_\d $ are material constants. 
{In this work, the stress norm $Y$ is computed as
	\begin{equation} \label{eq:dpy}
		Y = A\,Y_s + B\,Y_p + C,
	\end{equation}
	where
	$Y_s = \sqrt{3\,\up{tr}(\up{dev}{\boldsymbol{\Sigma}}\,\up{dev}{\boldsymbol{\Sigma}})/2}$, 
	with $\up{dev}\boldsymbol{\Sigma} = \boldsymbol{\Sigma} - 
	(\up{tr}\boldsymbol{\Sigma}/3)\boldsymbol{I}$,
	is the von Mises stress and 
	$Y_p=\up{tr}\boldsymbol{\Sigma}/3$ accounts for the spherical part of the
	stress tensor. The choice $A = 1$, $B = C = 0$, gives $Y = Y_s$, that is, 
	the von Mises stress, while other choices of coefficients in Eq.~\eqref{eq:dpy} 
	are intended to describe a Drucker--Prager-type yield criterion.}

Note that to treat the damaged state as a 
Newtonian fluid, it is sufficient to take $ \tau_\d = \up{const} \ll 1 $, see 
Sec.\,\ref{sec:results}\ref{sec.convection} 
or \cite{DumbserPeshkov2016}. Non-Newtonian rheologies can be 
also considered if the proper function for $ \tau_\d(Y) $ is provided, e.g. see 
\cite{Jackson2019a}. 
Thus, the function $ \theta_1 = \tau_1 c_s(\xi)^2/3 |\AA|^{-5/3}$ is taken in such a way as to 
recover the Navier-Stokes 
stress tensor with the effective shear viscosity $ \eta = \frac16\rho_0\tau_1c_s^2 $ 
in the limit $ \tau_1 \ll 1 
$ \cite{DumbserPeshkov2016} {and is used for modeling of a natural convection problem in 
	Sec.\,\ref{sec:results}\ref{sec.convection}}. A pure elastic response of the intact material, 
	{as used as fault host rock in Sec. \,\ref{sec:results}\ref{sec.dr}} 
cases (i) and (ii), corresponds to $ \tau_\i =\infty.$
{By this means, all numerical examples presented throughout Sec.\ref{sec:results} follow the 
rheological formulation given by $\theta_1$ with varying parametrisation.}

The transition from the intact to the 
fully damaged state is governed by the damage variable $\xi \in [0,1]$ satisfying the 
kinetic-type equation \eqref{eqn.xi}, $\dot{\xi} =-\theta E_\xi$, with the source term depending on 
the state parameters of the 
medium (pressure, stress and temperature). In particular, the rate of damage $ \theta $  
is defined as 
\begin{equation}\label{theta}
	\theta = \theta_0 (1-\xi)(\xi+\xi_\epsilon) 
	\left[(1-\xi)\left(Y/Y_0\right)^a+\xi
	\left(Y/Y_1\right)\right] ,
\end{equation} 
where $\xi_\epsilon$, $Y_0$ and $Y_1,a$ are constants. 
$\xi_\epsilon$ is usually set to $\xi_\epsilon=10^{-16}$ in order to trigger the growth of $\xi$ 
with the 
initial data  $\xi=0$. 
We note that similar to the chemical kinetics, the constitutive functions of the damage process 
drive the system towards an equilibrium that is not simply defined as $ E_\xi = 0 $, but as $ 
\theta 
E_\xi = 0 $, e.g. see \cite{PKG-Book2018}.
As a result, the overall response of the material subject to damage (i.e., its stress-strain 
relation, see also \cite{GPRCrack})
is defined by the interplay of both irreversible processes; (i) the degradation of the elastic 
moduli controlled by \eqref{theta} 
and (ii) the inelastic processes in the intact and 
damaged phases controlled by \eqref{tau} and \eqref{taus}.
{In the numerical experiments carried out in {Sec.\,\ref{sec:results}\ref{sec:brazilian}}, 
	the damage kinetics $\xi$ also strongly couple with strain relaxation effects, by means of Eq. 
	\eqref{tau}.}
The function $\theta_2$, governing the rate of the heat flux relaxation, is taken as 
$\theta_2(\tau_2) = \tau_2 \frac{c_h^2}{\rho T}$ that yields the classical Fourier law of heat 
conduction with the thermal conductivity coefficient $\kappa=\tau_2 c_h^2 $ in the stiff relaxation 
limit ($\tau_2 \rightarrow 0$), see \cite{DumbserPeshkov2016}. For simplicity, the thermal 
parameters
of the intact and damaged phases are here assumed identical.

{
	Finally, we remark that the problem of parameter selection for our unified model of continuum 
	mechanics, 
	is a nontrivial task. Due to the large amounts of parameters, 
	the problem may need to be solved monolithically via numerical optimisation algorithms 
	applied to data obtained from observational benchmarks such as triaxial loading experiments.
	Nonetheless, in certain limiting cases, some rationale can be developed in order to 
	estimate parameters without empirically considering several trial choices.
	For example, brittle materials can be constructed by choosing a very high value for the 
	exponent $a$ 
	in Eq. \eqref{taus}. By this means, the rate of growth $\theta$ of the damage variable $\xi$ 
	will
	activate as a switch when $Y$ reaches the $Y_0$ threshold. In this specific case, $Y_0$ 
	can be chosen equivalently to a yield stress.
	Also, the sensitivity to tensile stresses can be modelled by resorting 
	to techniques that are routinely used in science and engineering, e.g., using the 
	Drucker--Prager 
	yield criterion to compute $Y$.
	In the Brazilian tensile fracture example in Sec.\,\ref{sec:results}\ref{sec:brazilian}, 
	$\beta_{\i,\d}$ are set to zero as the complex stress-dependent 
	mechanisms they control are not necessary for achieving the desired material behaviour.
	Controlling the relaxation time of the damaged state ($\tau_\d$)
	can be useful for modelling friction within a natural fault zone: if a very low 
	relaxation time is chosen, which can be easily achieved by setting 
	$\tau_{\d0} = 10^{-6}\up{s}$, $\alpha_\d = \beta_\d = 0$, the fault will exert no 
	tangential stresses
	on the surrounding intact rock, as if it were filled with an inviscid fluid.
	Specific frictional regimes and (time-dependent) plastic effects can be described 
	by properly choosing the relaxation times $\tau_{\i,\d}$ (via $\tau_{\i0,\d0}, 
	\alpha_{\i,\d}$, 
	$\beta_{\i,\d}$), 
	which in general may require more complex automatic optimisation strategies.}

\section{Numerical examples}\label{sec:results}

In this section we present a variety of numerical applications of the GPR model 
relevant for earthquake rupture and fault zones. The governing PDE system \eqref{eqn.GPR} is solved 
using the high performance computing toolkit \textit{ExaHyPE} \cite{Reinarz2020},
which employs an Arbitrary high order derivative (ADER) discontinuous Galerkin (DG) finite element 
method in combination with an \textit{a posteriori} subcell
finite volume limiter on space time adaptive Cartesian meshes (AMR). For details, the reader is 
referred to \cite{GPRCrack} and to  
\cite{Dumbser2014,Zanotti2015a,FrontierADERGPR,Axioms,DumbserPeshkov2016,Peano1,Peano2} and 
references therein.

\begin{figure}
	\centering	
	\includegraphics[width=0.99\textwidth]{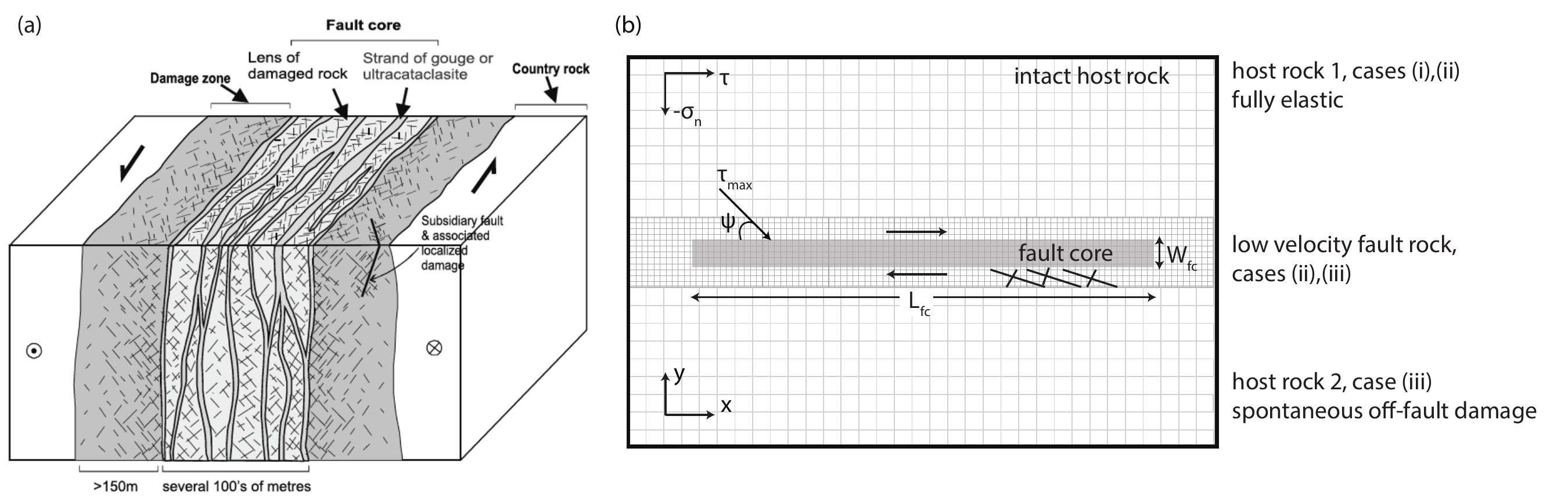}   
	\caption{ {(a) Typical strike-slip fault zone structure showing a multiple fault core 
	with associated damage zone in a quartzofeldspathic country rock (from 
			\cite{Mitchell2009}).}
		{(b) Sketch of the GPR model setup for 2D in-plane right-lateral shear fracture 
		under compression used throughout Sec.\ref{sec:results}\ref{sec.dr}.
			In light grey we depict the prescribed fault core of length $L_{fc}$ and width $W_{fc}$ 
			which is fully damaged ($\xi=1$) and embedded in intact host rock ($\xi=0$). 
			The material properties and rheology of the host rock and fault core differ and are 
			detailed in Tables S1 and S2.} 
		{Grey lines illustrate the initial mesh refinement, which can dynamically adapt as 
		detailed in Table S3.}
	}
	\label{fig.fault}
\end{figure}

\subsection{Earthquake shear fracture across a diffuse fault zone} \label{sec.dr}
%
%
{In the following}, we explore the GPR diffuse fault zone approach extending the modeling of 
dynamic 
earthquake rupture
beyond treatment as a discontinuity in the framework of elastodynamics.
{Fig. \ref{fig.fault} illustrates the model setup corresponding to the geological structure 
of a typical strike-slip fault zone.
	Dynamic rupture within the 'fault core' is governed by a friction-like behaviour achieved by 
	time-dependent modulation of 
	the shear relaxation time $\tau_\d$ of the fault core's fully damaged material. 
	At the onset of frictional yielding, 
	the shear relaxation time ($\tau_\d$) decreases exponentially as in \eqref{taus} with a 
	{time-dependent $\beta^{'}_\d$}. 
	The {temporal} evolution 
	of {$\beta^{'}_\d$} is modulated at a constant rate during rupture as 
	{$\beta^{'}_\d(t)\,=\,\beta_\d\,\min{\left(1,\, \max{\left(0,\, 
	1-C_1\,t\right)}\right)}$ 
		where $C_1$ and $\beta_\d$ are constant.}
	Visco-elastic slip accumulates across the diffuse fault core coupled to either fully elastic 
	wave propagation or Drucker-Prager type damage in the host rock.
}

\setlength{\belowcaptionskip}{-0.5cm}
\setlength{\abovecaptionskip}{-0.15cm} 

\textbf{i) Kinematic self-similar {Kostrov-like} crack.} 
We first model a kinematically driven \emph{non-singular self-similar shear crack} {analog 
to Kostrov's solution for a singular crack} \cite{Kostrov1964}
to study the relation between fault slip, slip rate and shear stress in comparison to traditional 
approaches, {while
	imposing tractions here avoids the full complexity of frictional rupture dynamics.}
The 2D setup \cite[e.g.]{Puente2009} assumes a homogeneous isotropic elastic medium ({Table 
S2}, $c_s=c_p/\sqrt{3}$), 
and a pre-assigned fault interface loaded by initial normal stress $\sigma_n=40\,\up{MPa}$
and shear stress $\tau=20\,\up{MPa}$. 
An in-plane right-lateral shear crack is driven by prescribing the (sliding) friction $\mu_f$ as 
linearly time-dependent weakening:
$\mu_f(x,t) = \text{max}\{ {f_d}, {f_s}-({f_s}-{f_d})(v_r 
t-|x|)/{R_c} \}$,
with {process zone size} ${R_c} = 250\,\up{m}$, rupture speed $v_r = 2000\,\up{m/s}$, 
static friction ${f_s} = 0.5$ and dynamic friction ${f_d} = 0.25$. 
%
We empirically find that choosing $C_1=10$ reproduces the propagating shear crack in the reference 
solution. 
Thus, $\beta^{'}_\d$ evolves linearly from $\beta_\d$ to 0 during rupture.  

%
\setlength{\belowcaptionskip}{-0.5cm}
\setlength{\abovecaptionskip}{-0.0cm} 
\begin{figure}[t!]
	\centering	
	\includegraphics[width=0.99\textwidth]{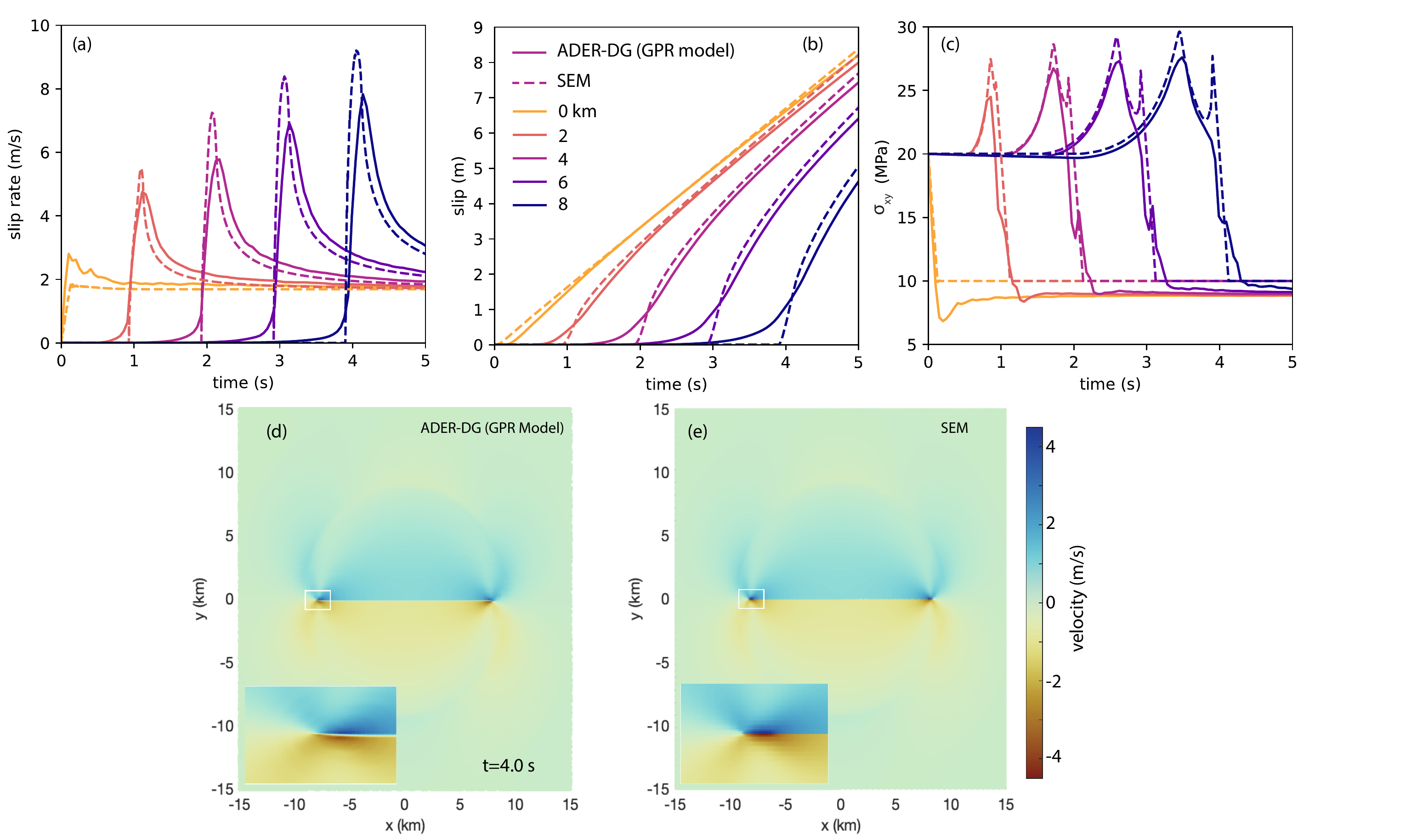}   
	\caption{Comparison of the self-similar {Kostrov-like} crack of the diffuse GPR model 
	(ADER-DG, $p=6$, {$W_{fc}=100\,\up{m}$, $L_{fc}=20\,\up{km}$,}  {fault core and 
	host rock material are 'host rock 1'}, static AMR) with the discrete fault spectral element 
	\textit{SEM2DPACK} ($p=6$, $h=100\,\up{m}$) solution;
		(a) Slip rate, (b) slip and (c) shear stress time series at increasing hypocentral 
		distances,
		(d,e) velocity wavefield at $t=4\,\up{s}$ (see also Animation S1), and zoom into the 
		rupture tip.}
	\label{fig.kostrov}
\end{figure}
We assume a fully-damaged {fault core} ($\xi=1$) of prescribed length $L_{fc}=20\,\up{km}$ 
and width $W_{fc}=100\,\up{m}$
embedded in a continuum material resembling intact elastic rock ($\xi=0$) as illustrated in 
Fig.\,\ref{fig.kostrov}a. Both, the fault {core} and the surrounding host rock are treated 
as the 
same continuum material besides their differences in $\xi$. 
The GPR specific material parameters are detailed as `host rock 1' (here, $\lambda_\d = \lambda_\i, 
\mu_\d=\mu_\i$) in Table S1 in the supplementary material. 
%
%
The model domain is of size $70\,\up{km}\times 70\,\up{km}$ bounded by Dirichlet boundary conditions
and employs a statically refined mesh surrounding the fault {core}.
The domain is discretised into hierarchical Cartesian computational grids, spaced $h=2800\,\up{m}$ 
at the coarsest level, and {$h=311\,\up{m}$} at the second refinement level ({Table 
S3}). 
We use polynomial degree $p=6$ 
and the subcell Finite Volume limiter counts $2\,p+1 = 13$ subcells in each spatial dimension. 
Fig.\ref{fig.kostrov}a-c compares slip, slip rate and shear traction during diffuse crack 
propagation 
modeled with the GPR model to a spectral element solution assuming a discrete fault interface 
spatially discretised with $h=100\,\up{m}$ with \textit{SEM2DPACK} \cite{sem2dpack}. 
%
The GPR model {analog} captures the kinematics (i.e., stress drop and fault slip) of the 
self-similar {singular} Kostrov crack as well as the emanated seismic waves 
(Fig.\,\ref{fig.kostrov}d,e and Animation S1), 
while introducing dynamic differences on the scale of the diffuse fault (zoom-in in 
Fig.\,\ref{fig.kostrov}d).
Slip velocity (Fig.\,\ref{fig.kostrov}a) remains limited in peak, 
similar to planar fault modeling with off-fault plastic deformation \cite{Gabriel2013}.
Fault slip (Fig.\,\ref{fig.kostrov}b) appears smeared out at its onset, yet asymptotically 
approaches the classical Kostrov crack solution.
Similarly, shear stresses (Fig.\,\ref{fig.kostrov}c) appear limited in peak and more diffuse, 
specifically with respect to the secondary peak associated with the passing rupture front. 
Importantly, (dynamic) stress drops are comparable to the expectation from fracture mechanics for a 
plane shear crack
(even though peak and dynamic level appear shifted).
At the crack tip, we observe an initial out-of-plane rotation within the fault {core} 
leading to a localised mismatch in the hypocentral region and at the onset of slip across the fault.
The GPR model approaches the analytical solution, as illustrated for increasing polynomial degree 
$p$ in the electronic supplement (Fig.\,S1).

\begin{figure}[t!]
	\centering
	\includegraphics[draft=false,width=0.93\textwidth]{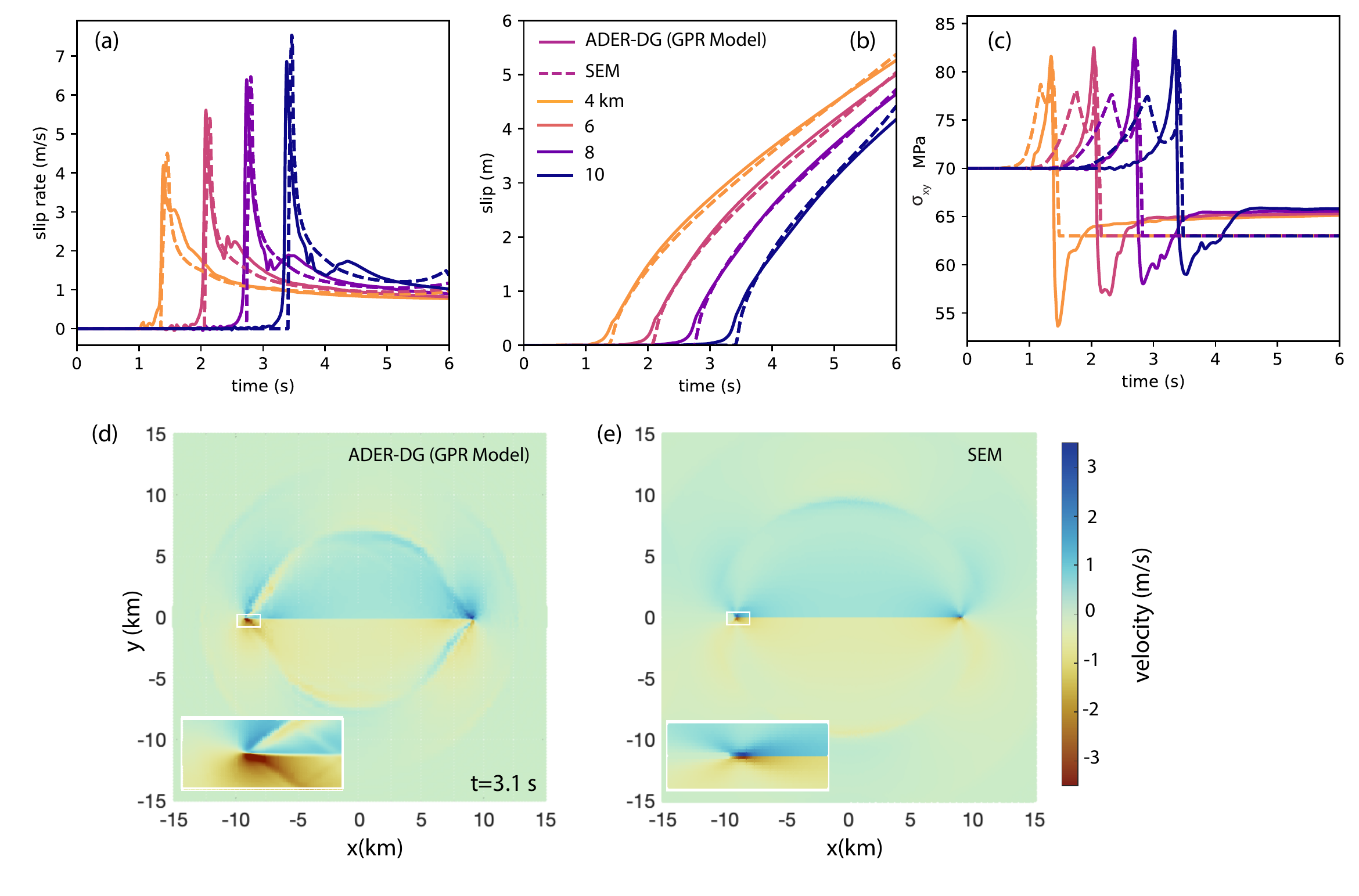}   
	\caption{Computational results for the 2D TPV3 dynamic rupture problem.
		Comparison of the diffuse interface GPR model (ADER-DG, $p=6$, 
		{$W_{fc}=100\,\up{m}$, $L_{fc}=30\,\up{km}$}, {fault core of 'low velocity 
		fault rock' embedded in 'host rock 1'}, static AMR) 
		with the discrete fault spectral element \textit{SEM2DPACK} solution ($p=6$, 
		$h=100\,\up{m}$), with (a) slip rate, (b) slip and 
		(c) shear stress time series at increasing hypocentral distance.
		(d,e) Radiated seismic wavefield in terms of particle velocity at $t=3.1\,\up{s}$ (see also 
		Animation S2). Zoom-in the crack tips highlights dynamic rupture complexity within the 
		{low-rigidity fault core}.} 
	\label{fig.tpv3}
\end{figure}

\textbf{ii) Spontaneous dynamic rupture.} 
We next model spontaneous dynamic earthquake rupture in a 2D version \cite{Puente2009} of the 
benchmark problem TPV3  \cite{Harris2018}
for elastic spontaneous rupture propagation defined by the Southern California Earthquake Center.
Our setup resembles the kinematic model (Fig.\,\ref{fig.fault}a) {including the 
time-dependent choice of $\beta^{'}_\d(t)$ with $C_1=10$} with an important distinction:
we assume a \emph{low-rigidity fault {core}} ({`low velocity fault rock'} in Table 
S1) by setting 
P-wave and S-wave velocity in the fault {core} 50\% lower, 
i.e. $\lambda(\xi)$ and $\mu(\xi)$ are decreased by 30\%, with respect to the intact rock. 
A 50\% reduction of seismic wave speeds matches natural fault zone observations. 
{The thickness of the low velocity fault rock unit equals the thickness of the fault core 
itself where $\xi=1$.} 
The surrounding country rock is again parameterised as fully elastic with the `host rock 1' GPR 
parametrisation ({Table S1}).
The fault {core} is $L_{fc}=30\,\up{km}$ long and $W_{fc}=100\,\up{m}$ wide, the domain size 
is $40\,\up{km}\times 40\,\up{km}$, 
initial loading is $\sigma_{yy}=-120\,\up{MPa}$ and $\sigma_{xy}=70\,\up{MPa}$. 
{The computational grid is spaced $h=1600\,\up{m}$ at the coarsest level, and 
$h=177\,\up{m}$ at the second refinement level (Table S3)}. 
Fig.\,\ref{fig.tpv3} compares, similar to the kinematic case, the diffuse low-rigidity fault 
ADER-DG GPR results to
an elastic discrete fault interface spectral element solution.
Fault slip rates (Fig.\,\ref{fig.tpv3}a) are limited in peak and are now clearly affected by 
smaller scale dynamic complexity, e.g. out-of-plane crack rotation and wave reflections, within the 
fault {core}. 
Fault slip (Fig.\,\ref{fig.tpv3}b) asymptotically resembles the discontinuous, elastic solution. 
Shear stresses (Fig.\,\ref{fig.tpv3}c) are smeared out and shifted, but capture (dynamic) stress 
drops, similar to the kinematic model in \textbf{i)}.
{We note that residual shear stress levels remain higher potentially reflecting oblique 
shear developing within the diffuse fault core and/or viscous behaviour within the fault core.}
The diffuse fault {core} slows down the emitted seismic waves, while amplifying sharp 
velocity pulses (Fig.\,\ref{fig.tpv3}d,e and Animation S2) aligning with observational findings 
\cite{Spudich2001}.
The GPR model successfully resembles frictional \emph{linear-slip weakening} 
behaviour\cite{Ida1972} 
within the fault {core} by defining:
$\mu_f(x,t) = \mathrm{max}\{ {f_d}, {f_s}-({f_s}-{f_d}) \delta(x,t) / 
D_c \}$,
with {slip-weakening distance} $D_c= 0.4$ m, ${f_s} = 0.677$ and ${f_d} = 
0.525$ similar to the discrete fault solution, 
$\delta(x,t)$ denotes here the diffuse slip within the fault {core} and is measured as the 
difference of displacements at its adjacent boundaries.  
%
%
%
Rupture is not initiated by an overstressed patch, which would be inconsistent with deforming 
material, 
but as a kinematically driven {Kostrov-like} shear-crack with $v_r=4000\,\up{m/s}$ and 
within a nucleation time of $t=0.5\,\up{s}$. 
{In the diffuse model, introducing the low velocity fault rock material within the fault 
core is required to limit rupture speed while resembling
	slip rate, slip and stress evolution of the discrete reference model. We conclude} that the 
	\emph{rheological fault core properties}, and not the friction law, 
control important crack dynamics such as rupture speed in our diffuse interface modeling, cf. 
\cite{Huang2014}.
A comparison of results assuming a further reduction of fault rock wave speeds to 60\% is discussed 
in the supplementary material. 
%

\begin{figure}
	\centering
	\includegraphics[draft=false,width=0.95\textwidth]{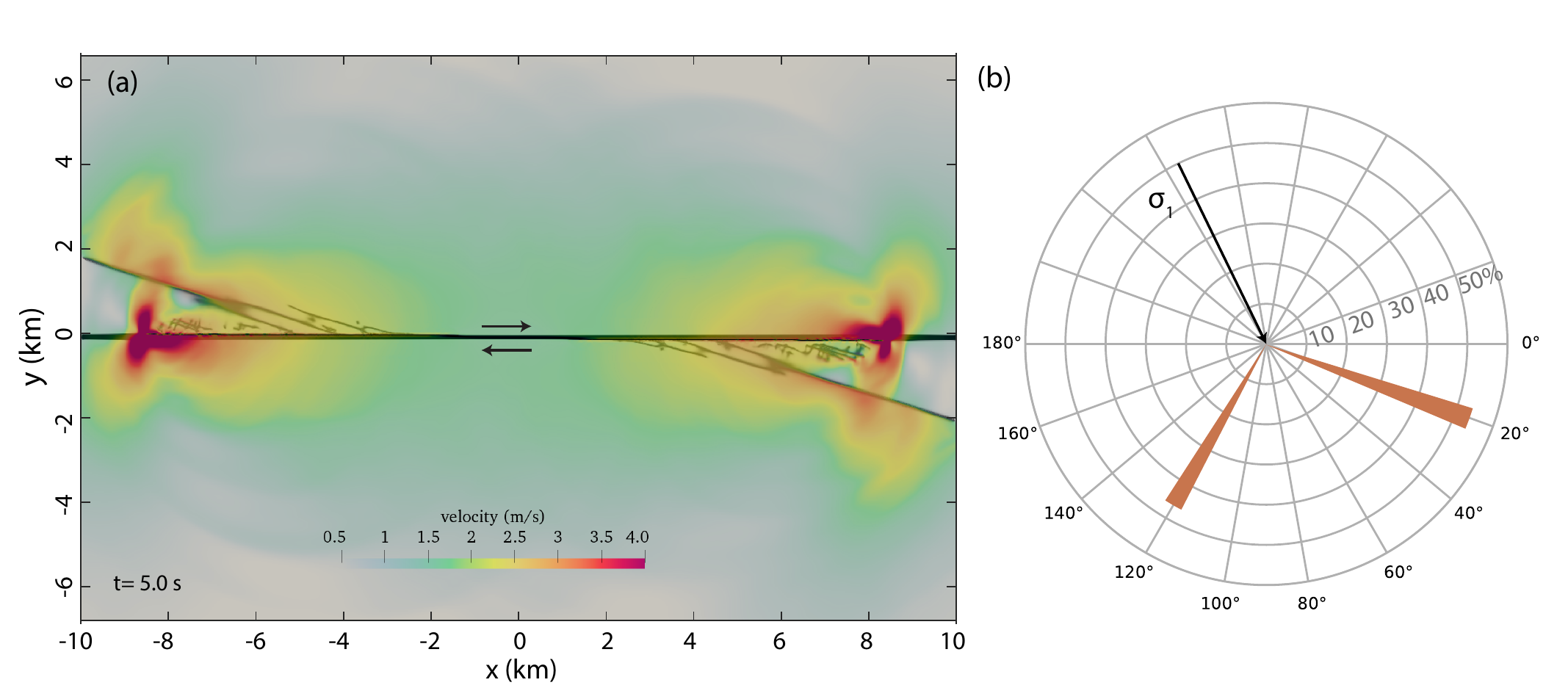}   
	\caption{Off-fault shear cracks spontaneously generated in the extensional quadrants of dynamic 
	earthquake rupture (TPV3) in the GPR model (ADER-DG,$p=6$, {$W_{fc}=100\,\up{m}$, 
	$L_{fc}=20\,\up{km}$}, {fault core of 'low velocity fault rock' embedded in 
	Drucker-Prager type 'host rock 2'}, dynamic AMR).
		(a) Velocity wavefield at $t=5.0\,\up{s}$ (see also Animation S3). Dark colours represent 
		the damage variable $\xi$ 
		illustrating the fault core initialized as fully damaged (cf. Fig.\,\ref{fig.kostrov}a) and 
		the propagating secondary off-fault cracks.
		(b) Polar diagram of the statistical orientation of off-fault shear cracks. The two 
		dominant orientations are  
		$\approx20^\circ$ and $\approx120^\circ$.  The maximum compressive stress ($\sigma_1$) has 
		an orientation angle of $65.3^\circ$.} 
	\label{fig.offfault}
\end{figure}

\textbf{iii) Dynamically generated off-fault shear cracks.} 
{Localized shear banding} is observed in the vicinity of natural faults spanning a wide 
spectrum of length scales \cite{Mitchell2009}, and contributes to the energy balance of earthquakes.
We model dynamically generated off-fault shear cracks by combining the spontaneous dynamic rupture 
model embedded in `low velocity fault rock' with `host rock 2' outside 
the fault core (Table S1,  $\mu_\d = 0.8571\,\mu_\i, \lambda_\d=\lambda_\i + 0.6667\,(\mu_\i - 
\mu_\d)$ in \eqref{Lame}). 'Host rock 2' is governed by
Drucker--Prager {yielding \cite{DruckerPrager1952,Templeton2008,Wollherr2018}} {as 
given by Eq.~\eqref{eq:dpy},
	with $A = 1/\sqrt{3}$, $B = \sin(\pi/18)$, and $C = -\cos(\pi/18)\cdot95\,\up{MPa}$}.
The model domain size is $20\,\up{km}\times 15\,\up{km}$ spatially discretised with $h=800\,\up{m}$ 
at the coarsest mesh level ({Table S3}).
We here use \emph{dynamic adaptive mesh refinement} (AMR) with two refinement levels and refinement 
factor $\mathfrak{r}=3$ to adapt 
resolution in regions where the material is close to yielding. 
The finest spatial discretisation is $h=89\,\up{m}$.
Fig.\,\ref{fig.offfault}a illustrates spontaneous shear-cracking in the \emph{extensional} 
quadrants of the main fault core, where the passing rupture induces a dynamic bimaterial effect 
\cite{Thomas2017}.
{While previous models \cite{Templeton2008} based on ideal plasticity without damage 
accumulation numerically capture
	the formation of single sets of shear bands in Drucker-Prager type off-fault material induced 
	by dynamic rupture propagation across a main fault,
	we here observe the formation of two conjugate
	sets of shear fractures:}
Cracks are distributed around two favourable orientations (Fig.\,\ref{fig.offfault}b).
{Spacing and length of these shear deformation bands \cite{MaElbanna2018,Okubo2019} may 
depend on GPR material parametrization ($Y_0,\beta_D$, cohesion, internal friction angle, etc., see 
Table S1 and \cite{GPRCrack}) as well as on the computational mesh and will motivate future 
analysis, as in Sec. \ref{sec:results}\ref{sec:brazilian}}.
High particle velocity is associated with the strong growth of off-fault shear stresses near the 
fault tip shifting from the propagation direction of the main rupture \cite{Freund1990}.
We observe the dynamic development of interlaced conjugate shear faulting (Animation S3) resembling 
recent high-resolution imaging of earthquakes \cite{Ross2019}.
%
%
%

\subsection{Crack formation in a rock-like disc}  \label{sec:brazilian}
\begin{figure}[!bp]
	\centering
	\includegraphics[draft=false,height=0.32\textwidth]{{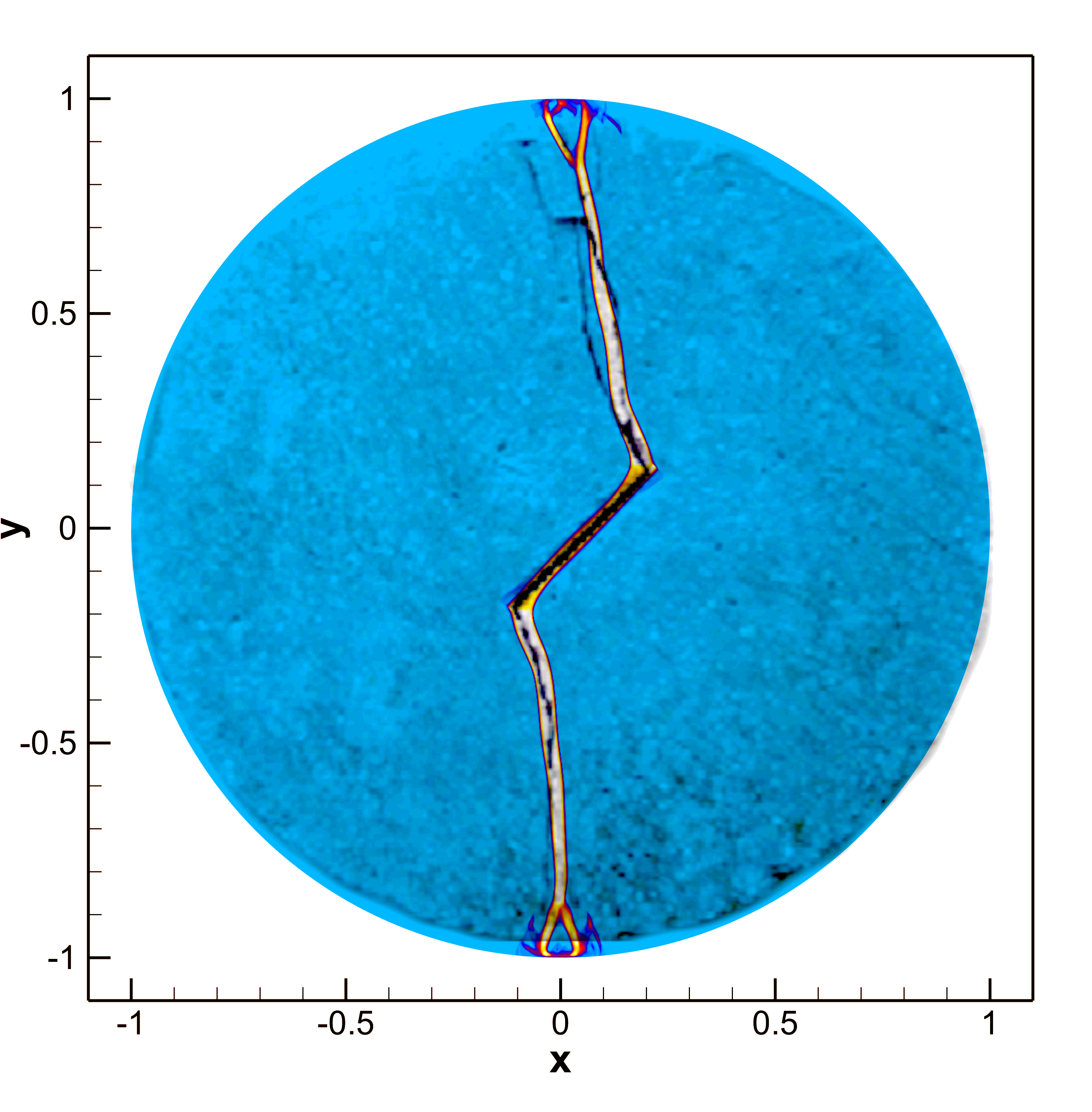}}%
	\includegraphics[draft=false,height=0.32\textwidth]{{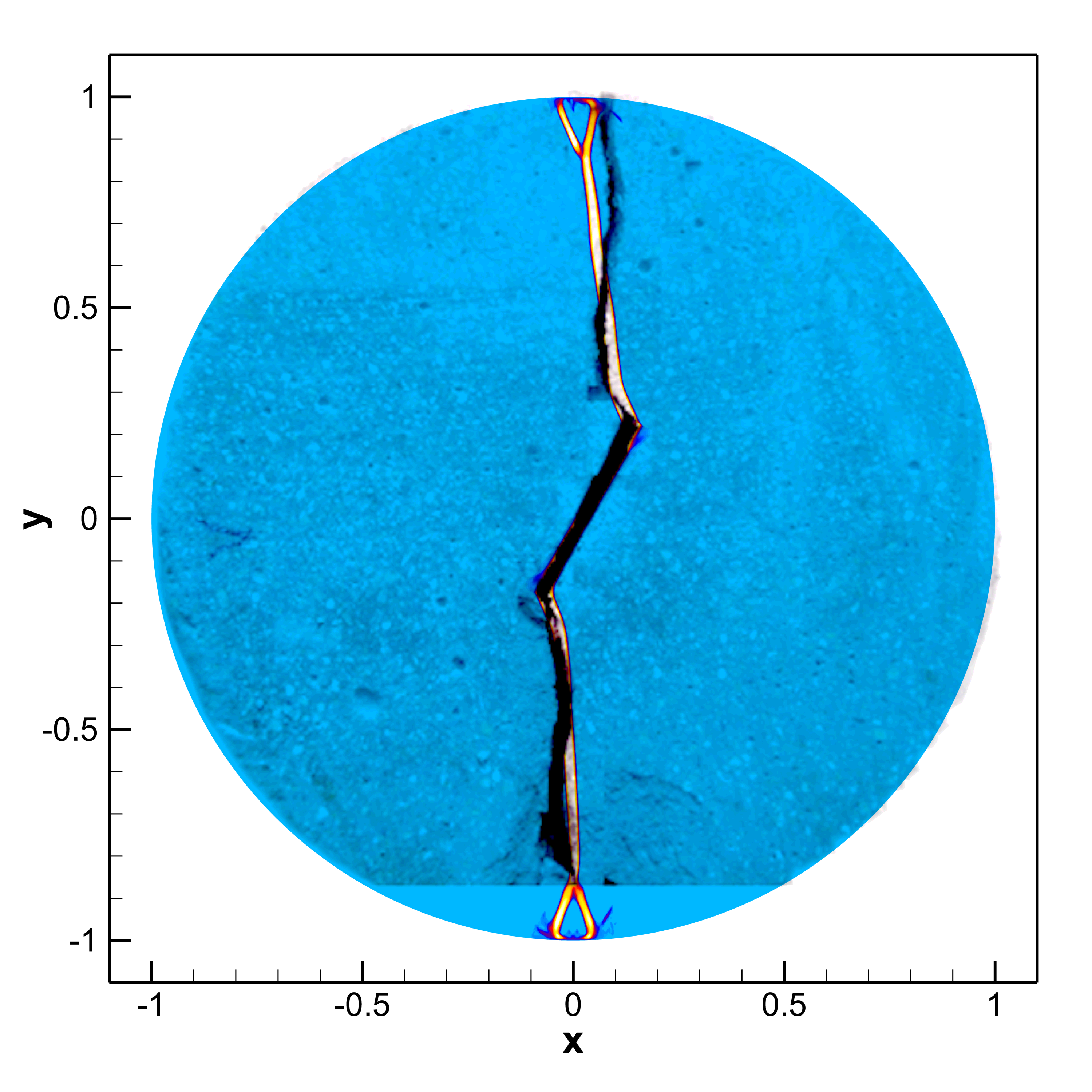}}%
	\includegraphics[draft=false,height=0.32\textwidth]{{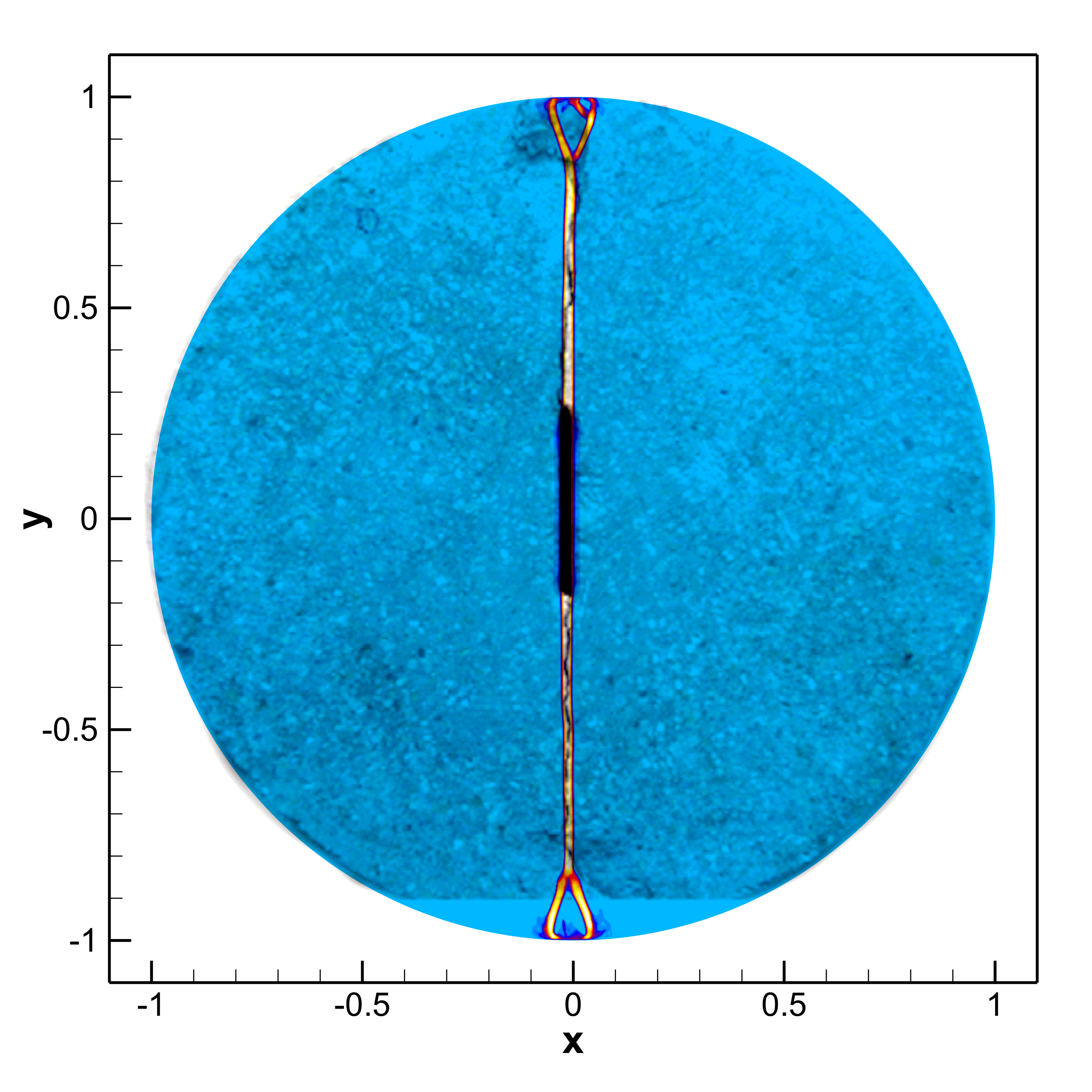}}%
	\includegraphics[draft=false,height=0.32\textwidth]{{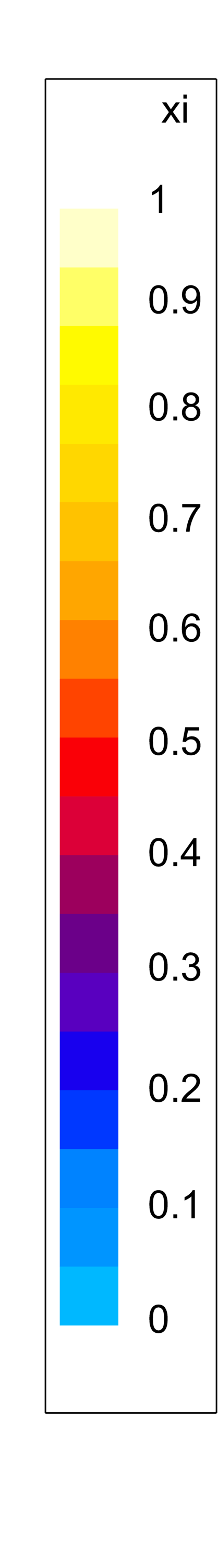}}%

	\caption{Crack formation in a rock-like disc under vertical load (Brazilian test) for different 
		angles of the pre-damaged area. Comparison of the contour colors of 
		the damage variable $\xi$ obtained in the numerical simulations of the GPR model with the 
		cracks 
		observed in experiments. {The simulation results are overlaid on top of the 
		photographs from 
			\cite{braziliandisc}.}
		From left to right: $45^\circ$, $60^\circ$ and $90^\circ$. Only the regions of the disc 
		where $\alpha>0.5$ are shown. }  
	\label{fig.disc}
\end{figure}

The GPR framework can be applied to capture tensile fracture, important for earthquake nucleation 
processes and the microscale of fault zone fracture and damage. 
We now show that our model is able to correctly describe the fracture mechanisms 
observed in laboratory settings. Specifically, we reproduce the experimental results of 
\cite{braziliandisc}
which involve the compression of a rock disc along its diameter (a so-called Brazilian test).
In this case the disc presents a central slit with a given orientation, 
which influences the
early stages of the failure of the rock sample. 
The test is carried out in two space dimensions on a square computational 
domain centered at the origin and with sidelength 2.2 units. The interface
of the disc is defined by setting $\alpha = 0$ outside of the unit-radius 
circle, without requiring 
a boundary-fitted mesh.
%
The material used in this test has been derived as a weakened variant of a granite-like brittle 
rock.
In particular, it replicates the strong difference in shear resistance found under compression or 
{tensile} loads.
The material is characterised by the following choice of parameters:
$\rho = 2620\,\up{kg/m^{3}}$,
$\mu_\i = \lambda_\i = 21.44\,\up{GPa}$, 
$\mu_\d = \lambda_\d = 150.08\,\up{MPa}$, 
$\theta_0 = 1$, 
$Y_0 = 10\,\up{MPa}$, 
$Y_1 = 1\,\up{Pa}$, 
$a = 60$,
$\tau_{\i0} = 10^5\,\up{s}$,
$\tau_{\d0} = 10^{-3}\,\up{s}$,
$\beta_\i = \beta_\d = 0$. For $|y| > 1$ the material is modified by setting $Y_0 = Y_1 = 
100\,\up{TPa}$ to model unbreakable clamps. 
Thermal effects are neglected.
For this test, the coefficients of the Drucker--Prager 
equivalent stress formula~\eqref{eq:dpy} are
$A = 1.0,\ B = 1.5,$ and $C = -2.0\,\up{MPa}$.
In Fig.\,\ref{fig.disc} we report the computational results from an ADER-DG ($p=3$) scheme on a 
uniform Cartesian mesh
of 192 by 192 cells, showing good agreement with the experimental data. For a detailed mesh 
refinement study, see the supplementary material. 

\setlength{\abovecaptionskip}{0.25cm} 

\subsection{Phase transition and natural convection in molten rock-like material} 
\label{sec.convection}
Seismic fault slip velocities and low thermal conductivity of rock can lead to the formation of 
veins of molten rock (pseudotachylytes), which are thought of as an unambiguous indicator of 
earthquake deformation, however, are not common features of active faults \cite{Sibson1975}.
In our model, the phase transition between solid and liquid occurs simply via the definition of the 
total energy by adding the contribution of the latent heat for $T > T_c$, see \eqref{E1a}, and by 
modifying the relaxation time for $T > T_c$.  
More precisely, {in this example, the relaxation time $ \tau_1 $ is not computed according 
	to \eqref{tau} and \eqref{taus} but is considered constant (time-independent) in the solid 
	state and is computed in 
	terms of the dynamic viscosity $\eta$ as 
	$\tau_1 = \frac{6 \eta}{\rho_0 c_s^2}$ for the molten state ($T > T_c$) treated as a Newtonian 
	fluid. Also, in this example, $ \theta_1 $ has to be taken as 
	$ \theta_1 = \tau_1 c_s(\xi)^2/3 |\AA|^{-5/3}$}, 
see the result of the asymptotic analysis presented in \cite{DumbserPeshkov2016}. 
In the supplementary material of this paper we validate our simple approach for phase transition 
for a well-known benchmark problem with exact solution, namely the Stefan problem, see 
\cite{baehr_en}. The obtained results clearly show that the proposed model can properly deal with 
heat conduction and phase transition between liquid and solid phases.



\begin{figure}
	\centering
	\begin{tabular}{cc}
		\includegraphics[width=0.45\textwidth]{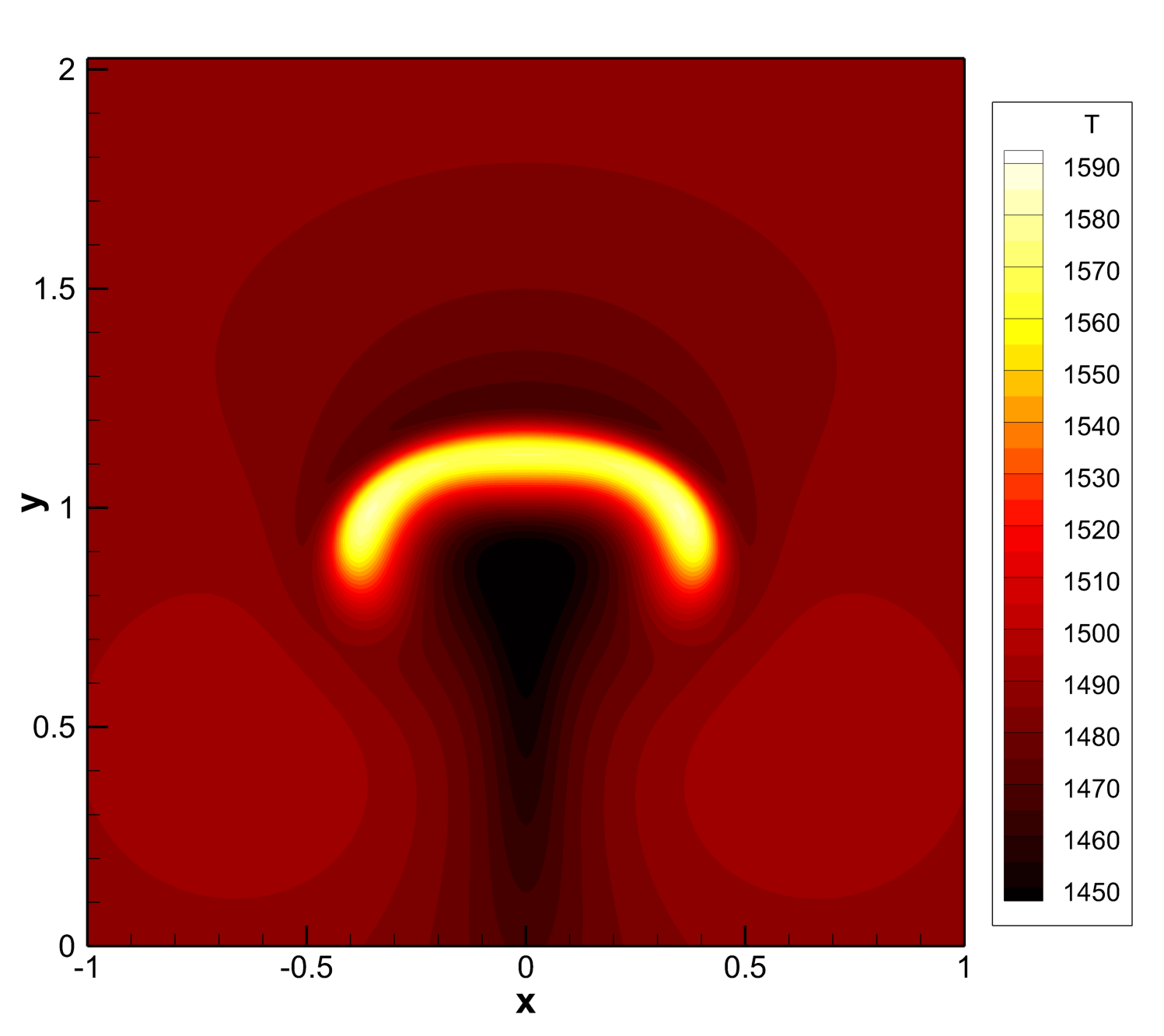} &  
		\includegraphics[width=0.45\textwidth]{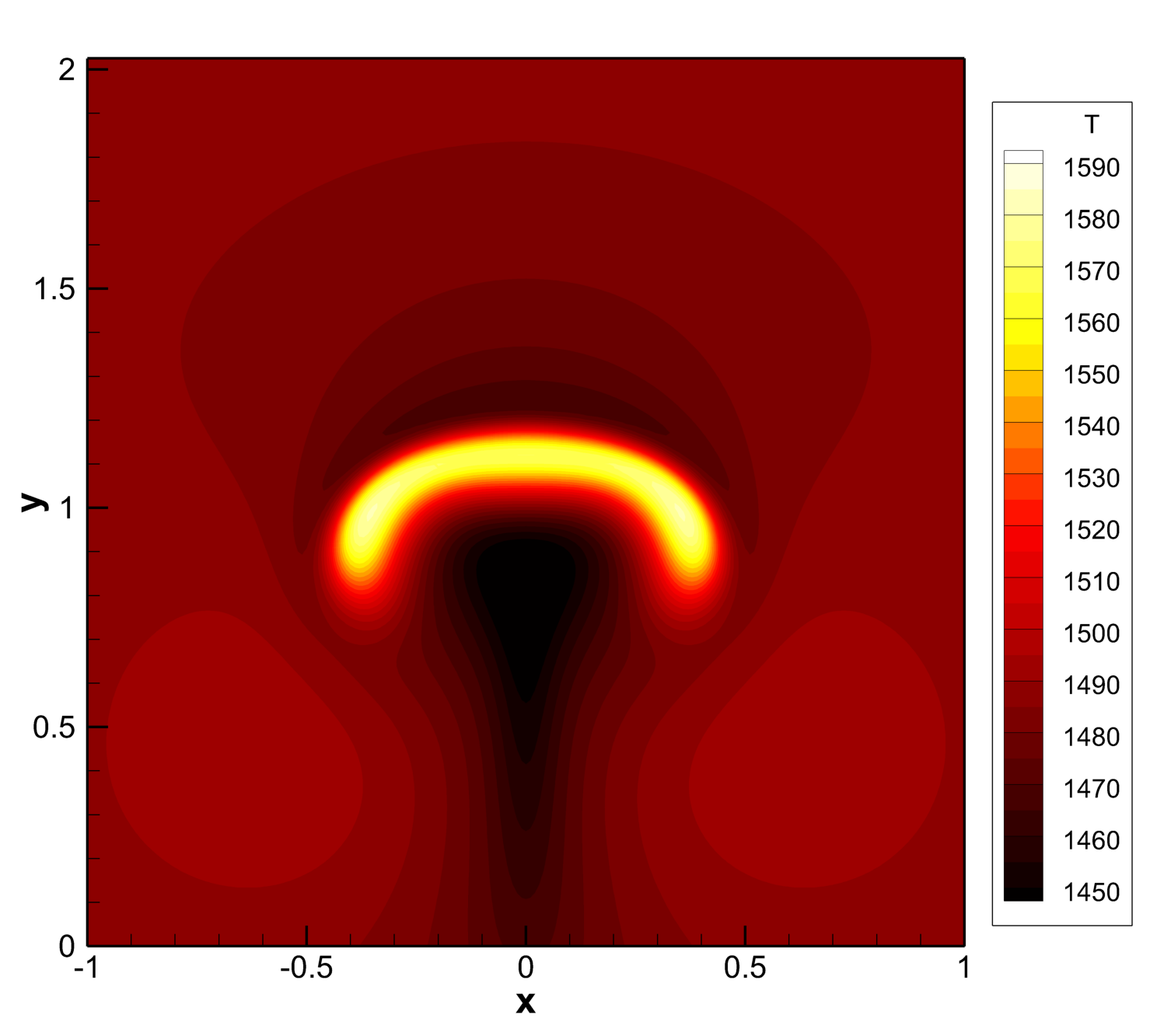}
	\end{tabular} 
	\caption{Temperature contours for the rising bubble problem in molten rock-like material at 
	time $t=4$. Solution obtained with the GPR model (left) and Navier-Stokes reference solution 
	(right). The melting temperature is set to $T_c=1000$. }
	\label{fig.bubble}
\end{figure}

Next, we show the capability of the GPR model to describe also the motion of viscous fluids under 
the influence of gravity. 
The stresses acting on faults are key initial conditions for earthquakes and seismic fault 
dynamics, but are poorly known. At very long time scales, these initial conditions are governed by 
plate tectonics and mantle convection, which is included in the GPR model as a special case. 
We therefore simulate a rising bubble in molten rock-like material. In the following, we use SI 
units.
The critical temperature is set to $T_c=1000$, the latent heat is $h_c = 1000$, the gravity vector 
is 
$\mathbf{g}=(0,-9.81)$ and the dynamic viscosity of the molten material is $\eta=20$. 
We furthermore set the remaining parameters to $\rho_0 = 2000$, $\gamma=2$, 
$p_0 = 2 \cdot 10^5$, $c_v = 0.1$, $c_s=5$, $\alpha=5$ and $\lambda = 0.2$. 
Initially we set $T=1500$, $v_i=0$, $\AA = \II$, $\JJ=0$, $p=10^5 - \|\mathbf{g}\| \rho_0 y$ 
and a hot circular bubble of radius $R=0.2$ is initially centered at $\mathbf{x}_c = 
(0,0)$ with a temperature increase of $\Delta T = 200$ for 
$\left\| \mathbf{x} - \mathbf{x}_c \right\| \leq R$.  The domain is $\Omega = [-2,2] \times [-1,3]$ 
and simulations are carried 
out until $t=4$ with an ADER-DG ($p=3$) scheme on a mesh of $200 \times 200$ elements. For 
comparison, we run two simulations, one with the GPR model presented in this paper and another 
simulation with the compressible Navier-Stokes equations, which serves as a reference solution for 
the GPR model in the viscous fluid limit. The 
computational results are depicted in Fig. \ref{fig.bubble}, where we can observe an excellent 
agreement. This demonstrates that the model presented in this paper is able to correctly describe 
also natural convection in molten material when $T>T_c$.  
\section{Summary and Outlook}
We have presented a unified hyperbolic model of inelasticity that incorporates finite strain 
elastoviscoplasticity and viscous fluids in a single PDE system,
coupled with a hyperbolic model for continuous modeling of damage, including brittle and 
ductile fracture as particular cases. 
The governing equations are formulated in the Eulerian frame and via a diffuse interface approach 
permit arbitrary geometries of fractures and material boundaries without the necessity of 
generating interface-aligned meshes.
%
{We emphasize that the presented \textit{diffuse interface} approach is not merely a way to 
	regularize otherwise singular problems as posed by earthquake shear crack nucleation and 
	propagation along zero-thickness
	interfaces, but potentially allows to fully model volumetric fault zone shearing during
	earthquake rupture, which includes spontaneous partition of fault slip into intensely
	localized shear deformation within weaker (possibly cohesionless/ultracataclastic) fault-core 
	gouge and
	more distributed damage within fault rocks and foliated gouges.}
The model capabilities were demonstrated in several 2D examples related to rupture processes in 
earthquake fault zones.
We compare kinematic, fully dynamic and off-fault damage GPR diffuse rupture to models employing 
the traditional elasto-dynamic viewpoint of a fault, namely a planar surface across which slip 
occurs.
We show that the continuum model can reproduce and extend classical solutions, 
while introducing dynamic differences (i) on the scale of pre-damaged/low-rigidity fault zone, 
such as out-of-plane rupture rotation, limiting peak slip rates, non-frictional control of rupture 
speed;
and (ii) on the scale of the intact host rock, such as conjugate shear cracking in tensile lobes 
and amplification of velocity pulses in the emitted wavefield. 
A natural next step is to combine the successful micro fracture laboratory scale Brazilian tests 
with dynamic rupture to span the entire scales of fault zone fracture. The GPR parameters for the 
host rock and fault zone rock material can also be calibrated to resemble natural rock, as e.g. 
Westerly granite \cite{Lockner1998}.
Also, using the computational capabilities of the model's \textit{ExaHyPE} implementation, one can 
study related effects on ground shaking (see \cite{GPRCrack, Tavelli2019} for GPR modeling of 3D 
seismic wave propagation with complex topography) and detailed 3D fault zone models 
{\cite[cf.]{Wollherr2019,Ulrich2019,Ulrich2019b}} including trapped/head waves interacting 
with dynamic rupture \cite{Huang2014}.
%
Inelastic bulk processes are important during earthquake rupture \cite[e.g.,]{Ulrich2020}, but also 
in between seismic 
events, including off-fault damage and its healing, dynamic shear localization and interseismic 
delocalization, and visco-elasto-plastic relaxation. 
Since the unified mathematical formulation proposed in this paper is able to describe 
elasto-plastic solids as well as viscous fluids, future work will also concern the study of fully 
coupled models of dynamic rupture processes triggered by mantle convection and plate tectonics. 
Extensions to non-Newtonian fluids will be considered, {as well as to elasto-plastic 
	saturated porous media}, see e.g. the recent work presented in 
	\cite{Jackson2019a,Poroelast2020}.  
We also plan more detailed investigations concerning the onset of melting 
processes in shear cracks. Finally, we note that the material failure is due to the accumulation of 
microscopic defects (micro-cracks in rocks or dislocations in crystalline solids). 
It is thus interesting to remark that the distortion field being the field of non-holonomic 
basis triads provides a natural basis for 
further development of the model towards a micro-defects-based damage theory. 
This can be achieved via concepts of the Riemann-Cartan geometry, such 
as torsion discussed in \cite{PRD-Torsion2019}.


\paragraph{Data Accessibility.} \textit{ExaHyPE} is free software hosted at \url{www.exahype.org}. 
The presented numerical examples will be accessible and reproducible at 
\url{https://gitlab.lrz.de/exahype/ExaHyPE-Engine} and \url{https://github.com/TEAR-ERC/GPR2DR}. 
%
%
%
\paragraph{Funding.} This research has been supported by the European Union's Horizon 2020 Research 
and Innovation Programme under the projects \textit{ExaHyPE}, grant no. 671698, ChEESE, grant no. 
823844 and TEAR, grant no. 852992. 
MD and IP also acknowledge funding from the Italian Ministry of Education, University 
and Research (MIUR) via the Departments of Excellence Initiative 2018--2022  
attributed to DICAM of the University of Trento (grant L. 232/2016) and the PRIN 2017 project 
\textit{Innovative numerical methods for evolutionary partial differential equations and  
applications}. 
SC was also funded by the Deutsche Forschungsgemeinschaft (DFG) under the project DROPIT, grant no. 
GRK 2160/1. 
ER was also funded within the framework of the state contract of the Sobolev Institute of 
Mathematics (project no. 0314-2019-0012).  
AG also acknowledges funding by the German Research Foundation (DFG) (grants no. GA 2465/2-1, GA 
2465/3-1), by KAUST-CRG (grant no. ORS-2017-CRG6 3389.02) and by KONWIHR (project NewWave).
Computing resources were provided by the Institute of Geophysics of LMU Munich \cite{oeser2006} 
and the Leibniz Supercomputing Centre (project no. pr63qo).

\paragraph{In memoriam.} This paper is dedicated to the memory of Anne-Katrin Gabriel (*March 7, 
1957 \textdagger July 25, 2020) whose creativity and curious mind will live on -- in science with 
her daughter Alice.


\bibliographystyle{plain}
\bibliography{biblio}

\pagebreak
\begin{center}
	\textbf{\large Supplementary Material for:\ \ } \textbf{\Large A unified first order hyperbolic 
	model for nonlinear dynamic rupture processes in diffuse fracture zones}
\end{center}
\setcounter{equation}{0}
\setcounter{figure}{0}
\setcounter{table}{0}
\setcounter{page}{1}
\setcounter{section}{0}
\makeatletter
\renewcommand{\theequation}{S\arabic{equation}}
\renewcommand{\thefigure}{S\arabic{figure}}
\renewcommand{\thetable}{S\arabic{table}}

\newcommand\smallsection{%
	\titleformat{\section}
	{\bfseries\large}{\thesection}{1em}{}
}

\section{GPR material parametrisation, geometry and computational mesh of the numerical examples 
of section \textit{3(a) Earthquake shear fracture across a diffuse fault zone}}

\begin{table}[th!]   
\begin{center}
	\begin{tabular}{c | c c c c c c c c c c}
		material  & $Y_0$ (Pa)  & $Y_1$ (Pa) &  $a$ & $\alpha_\i$ & $\beta_\i$	 & $\alpha_\d$ & 
		$\beta_\d$ & 
		$\mu_{\i,\d}$ & $\lambda_{\i,\d}$	\\
		\hline
		host rock 1& 1.8e22  & 1.0e20 &  32.5  & 36.25 & 0.0	 & 36.25 & 1.0e-6 & 1.0 & 0.0	\\
		host rock 2& 1.8e8 & 1.0e10 &  32.5  & 36.25 & 0.0	 & 36.25 & 1.0e-6  &  0.8571  & 0.6667	
		\\
		low-velocity fault rock &  1.8e22  &  1.0e20  &  32.5  &  36.25 & 0.0	 & 12.687 & 1.75e-7 
		& 1.0 & 
		0.0	\\
	\end{tabular}
	\caption{GPR material parameterisation for earthquake shear rupture models in section 3(a). 
	Host rock 1 is the fully elastic material used in the full model domain of case (i), the 
	kinematic Kostrov-like crack. The fully elastic \textbf{host rock 1} is also surrounding the 
	fault core 
	in case (ii), the spontaneous dynamic rup-ture example. \textbf{Host rock 2} is the 
	Drucker-Prager type 
	off-fault material of case (iii) in which off-fault shear cracks nucleate and propagate. The 
	(visco-)elastic \textbf{low-velocity fault rock} material is used within the fault core in 
	cases (ii) 
	and (iii) and corresponds to the fully elastic host rock 1 with 50\% decreased seismic 
	velocities. Base relaxation times have been set as $ \tau_{\i0} = 10^{-6}\cdot\mu_\i $  and 
	$ \tau_{\d0} = 10^{-6}\cdot\mu_\d $ .
	}
	\label{table:1}
\end{center}
\end{table}

\begin{table}[th!]   
\begin{center}
	\begin{tabular}{c | c c c c c c c c c}
		case  &  $\rho$ (kg/m$ ^3 $)  & $c_p$ (m/s) &  $c_s$ (m/s) & $f_s$ & $f_d$	 & $D_c$ or $ 
		R_c $ (m) & $\sigma_{yy}$ (Pa) & $\sigma_{xy}$ (Pa) 	\\
		\hline
		(i)   & 2500 & 4000	& 2309 & 0.4 & 0.2 & 250 & -40e6 & 20e6	      \\
		(ii)  & 2676 & 6000 & 3464 & 0.677 & 0.525 & 0.4 & -120e6 & 70e6  \\
		(iii) & 2676 & 6000 & 3464 & 0.677 & 0.525 & 0.4 & -120e6 & 70e6  \\
	\end{tabular}
	\caption{Effective GPR material properties and reference friction parameters for section 3(a). 
	Case (i) specifies the parameters of the self-similar Kostrov-like crack model. Cases (ii) and 
	(iii) specify the parameters of the spontaneous dynamic rupture in fully elastic and 
	Drucker-Prager type plastic material. Note that the low-velocity fault rock material of the 
	fault core in cases (ii) and (iii) has $ c_p $ and $ c_s $ 50\% lower than host rock 1.
	}
	\label{table:2}
\end{center}
\end{table}

\begin{table}[th!]   
	\begin{center}
		\begin{tabular}{c | c c c c c c c c}
			case  &  \thead{domain \\ length (m)} & \thead{domain \\ width (m)} & \thead{fault core 
			\\ length $ L_{fc} $ (m)} &	\thead{fault core \\ width $ W_{fc} $ (m)} & 
			\thead{coarsest mesh \\ size h (m)} & \thead{static refinement \\ width 
			(m)} & \thead{mesh refinement \\ level; factor r} 	\\
			\hline
			(i)   & 70000 & 70000 & 20000 &	100 & 2800 & 1400 &	2; 3 \\
			(ii)  & 40000 & 40000 & 30000 & 100 & 1600 & 800  &	2; 3 \\
			(iii) & 20000 & 15000 & 20000 & 100 & 800  & 400  &	2; 3 \\
		\end{tabular}
		\caption{Geometry and computational mesh used in section 3(a). The domain is discretised 
		into hierarchical Cartesian computational grids of two mesh refinement levels and using 
		mesh refinement factor $ r=3 $. In case (i) it is spaced $ h=2800 $\,m at the coarsest 
		level, and 
		$ h_{min}=311 $\,m at the second refinement level in in the vicinity of the fault core, in 
		case (ii) 
		$ h=1600 $\,m and $ h_{min}=177 $\,m, respectively, and in case (iii), which uses dynamic 
		AMR, $ h=800 $\,m and 
		$ h_{min}=89 $\,m.
		}
		\label{table:3}
	\end{center}
\end{table}

\vspace{-0.5cm}
\section{Earthquake shear fracture animations}

\paragraph{Supplementary Animation S1:} A map-view animation of the p=6 GPR Kostrov-like crack and 
seismic wave radiation model (case i) in terms of velocity is provided as supplementary material 
and accessible here:\\
\url{https://youtu.be/CBcbLeqaB_A}

\vspace{-0.4cm}
\paragraph{Supplementary Animation S2:} A map-view animation of the p=6 GPR dynamic rupture and 
seismic wavefield TPV3 model (case ii) in terms of velocity is provided as supplementary material 
and accessible here: \\
\url{https://youtu.be/wx6-m6XS8C0}

\vspace{-0.4cm}
\paragraph{Supplementary Animation S3:} A map-view animation of the p=6 GPR co-seismic off-fault 
shear cracks spontaneously generated in the extensional quadrants of dynamic earthquake rupture 
TPV3 model and seismic wavefield (case iii) in terms of velocity is provided as supplementary 
material and accessible here:\\
\url{https://youtu.be/95oEDZBqIiE}

\section{P-refinement analysis for the GPR self-similar kinematic Kostrov crack model}
\vspace{0.5cm}
\begin{minipage}[t]{1.0\textwidth}
	\floatbox[{\capbeside\thisfloatsetup{capbesideposition={right,top},capbesidewidth=10cm}}]{figure}[\FBwidth]
	{\caption{Fault slip rate of the self-similar Kostrov-like crack (case i) modeled with the 
	diffuse 
	ADER-DG GPR method under varying polynomial degree p at 4 km hypocentral distance. As 
	refer-ence 
	the discrete fault spectral element SEM2DPACK solution is given.}\label{fig.s1}}
	{\includegraphics[trim= 0 0 0 30,width=6cm]{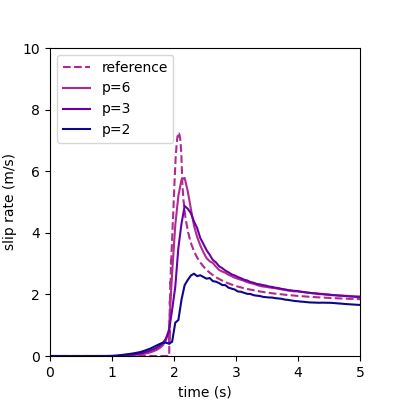}}
\end{minipage}

\section{Low-rigidity fault zone effects on spontaneous dynamic rupture (TPV3 model)}
\vspace{0.5cm}
\begin{minipage}[t]{1.0\textwidth}
	\floatbox[{\capbeside\thisfloatsetup{capbesideposition={right,top},capbesidewidth=5cm}}]{figure}[\FBwidth]
	{\caption{Further reduction of P- and S-wave velocity of the rheology of the ‘low-velocity 
	fault rock’ for GPR TPV3 (case ii) results. (a,c) are taken from Fig. 3 (a,c) in the main 
	manuscript. (b,d) illus-trate a decrease in rupture speed and increase in peak slip rate while 
	the stress drop is equivalent. All computational results are solving the 2D TPV3 dynamic 
	rupture problem of the diffuse interface GPR model using an ADER-DG ($ p = 6 $, static AMR) 
	scheme 
	with the discrete fault spectral element SEM2DPACK solution ($ p = 6 $, $ h = 100 $ m), with 
	(a,b) slip 
	rate and (c,d) shear stress measured at increasing hypocentral distance. }\label{fig.s2}}
	{\includegraphics[trim= 0 0 0 30,width=10cm]{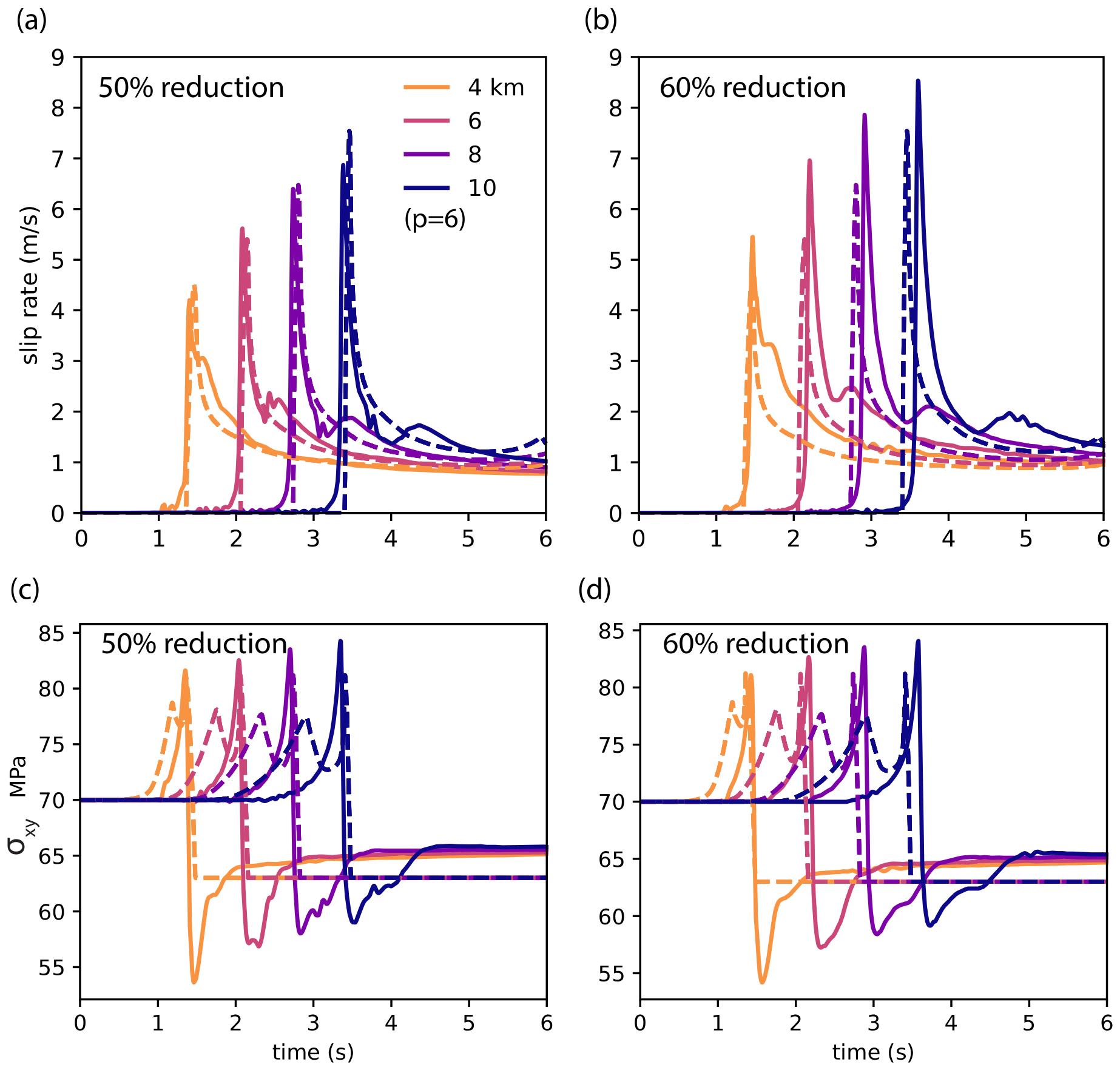}}	
\end{minipage}

\section{Validation of the phase-transition via the Stefan problem and comparison with the exact
	solution}

Here we validate this simple approach on a well-known benchmark problem with exact solution, 
namely the Stefan problem, see \cite{baehr_en}. We consider the computational domain $\Omega=[0,2] 
\times [-0.1, +0.1]$ with periodic boundaries in $y$ direction. The model parameters are set to 
$\gamma=2$, $p_0 = 10^7$, $\rho_0 = 1000$, $\alpha = 10^4$, $\lambda = 10^4$, $c_s=0$. The initial 
densities are chosen as $\rho_L = 2000$ for $x<0$ and $\rho_R = 1000$ for $x \geq 0$. Velocity, $ 
\JJ $ and the 
pressure are initially set to zero and $\AA$ is initialized with the identity matrix. The critical 
temperature is chosen as $T_c=900$ and the latent heat is $h_c = 10^4$. We now solve the GPR model
until time $ t = 0.2 $ with a fourth order ADER-DG scheme on a uniform Cartesian mesh of $ 400 
\times 4 $ elements
and compare our numerical results with the exact solution of the Stefan problem\footnote{H.D. Baehr 
and K. Stephan. Heat and Mass Transfer. Springer, 2011} at constant density,
see \Fig\,\ref{fig.stefan}. The agreement is good, in particular considering the fact that
in the GPR model density and pressure do not remain constant in time, as they are instead in the 
exact
solution of the Stefan problem. The obtained results clearly show that the proposed model can 
properly
deal with heat conduction and phase transition between liquid and solid phase

\vspace{0.5cm}
\begin{minipage}[t]{1.0\textwidth}
	\floatbox[{\capbeside\thisfloatsetup{capbesideposition={right,top},capbesidewidth=5cm}}]{figure}[\FBwidth]
	{\caption{Computational results obtained with the GPR model for the Stefan problem
			at time $t=0.2$. Comparison with the exact solution of the original Stefan problem at 
			constant density. The melting temperature is set to $T_c=900$. }\label{fig.stefan}}
	{\includegraphics[trim= 0 0 0 30,width=7cm]{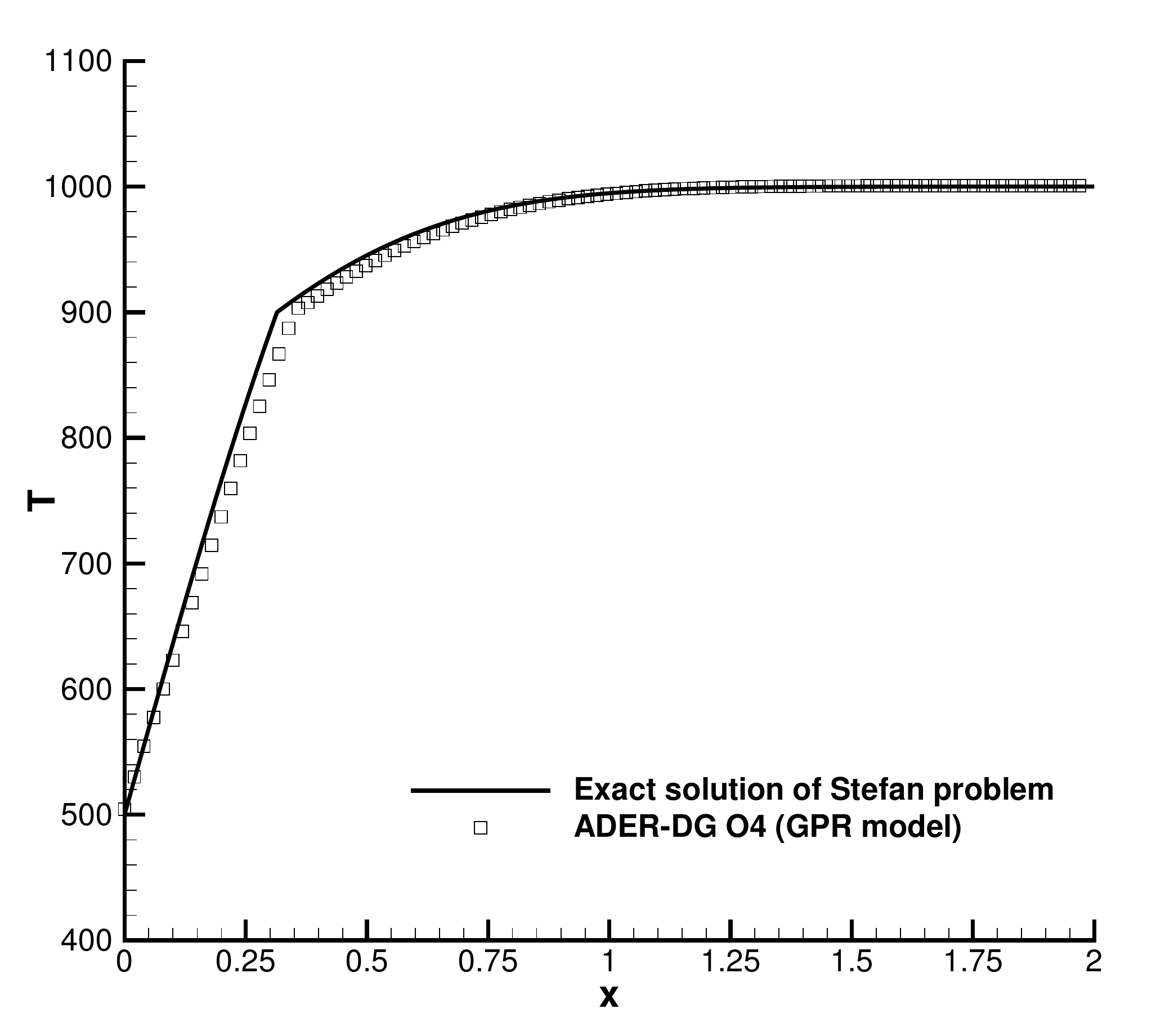}}	
\end{minipage}

\section{Mesh convergence for the Brazilian disc test}

Here we report the results of a mesh convergence study carried out on the Brazilian disc test. We 
aim
to show that the main directions in which cracks develop do not depend on the mesh resolution used 
in
the computations. The simulations are carried out with a fourth order ADER-DG scheme with 
MUSCL-Hancock Finite Volume subcell limiting.
We would like to stress that both the intensity of material damage and the problem geometry, as well
as the local material properties, are represented by means of scalar fields, so that arbitrarily 
complex
configurations can be simulated without requiring to use any ad-hoc meshing strategies. The setup 
for
the test problem is shown in \Fig\ref{fig.S4}, where we plot the fields associated 
with the
geometrical field $ \alpha $ (which identifies the presence of solid material or of air/vacuum) and 
with a 
material-
type indicator. The use of two different sets of material properties is due to the need to model a 
clamping
apparatus that we approximate as unbreakable. Note that \emph{no} feature jump coincides with a cell
boundary and in general the geometry can be skewed with respect to the computational grid, or even
curved.

\begin{minipage}[t]{1.0\textwidth}
	\floatbox[{\capbeside\thisfloatsetup{capbesideposition={right,top},capbesidewidth=5cm}}]{figure}[\FBwidth]
	{\caption{Computational results obtained with the GPR model for the Stefan problem
			at time $t=0.2$. Comparison with the exact solution of the original Stefan problem at 
			constant density. The melting temperature is set to $T_c=900$. }\label{fig.S4}}
	{	\includegraphics[draft=false,width=0.33\textwidth]{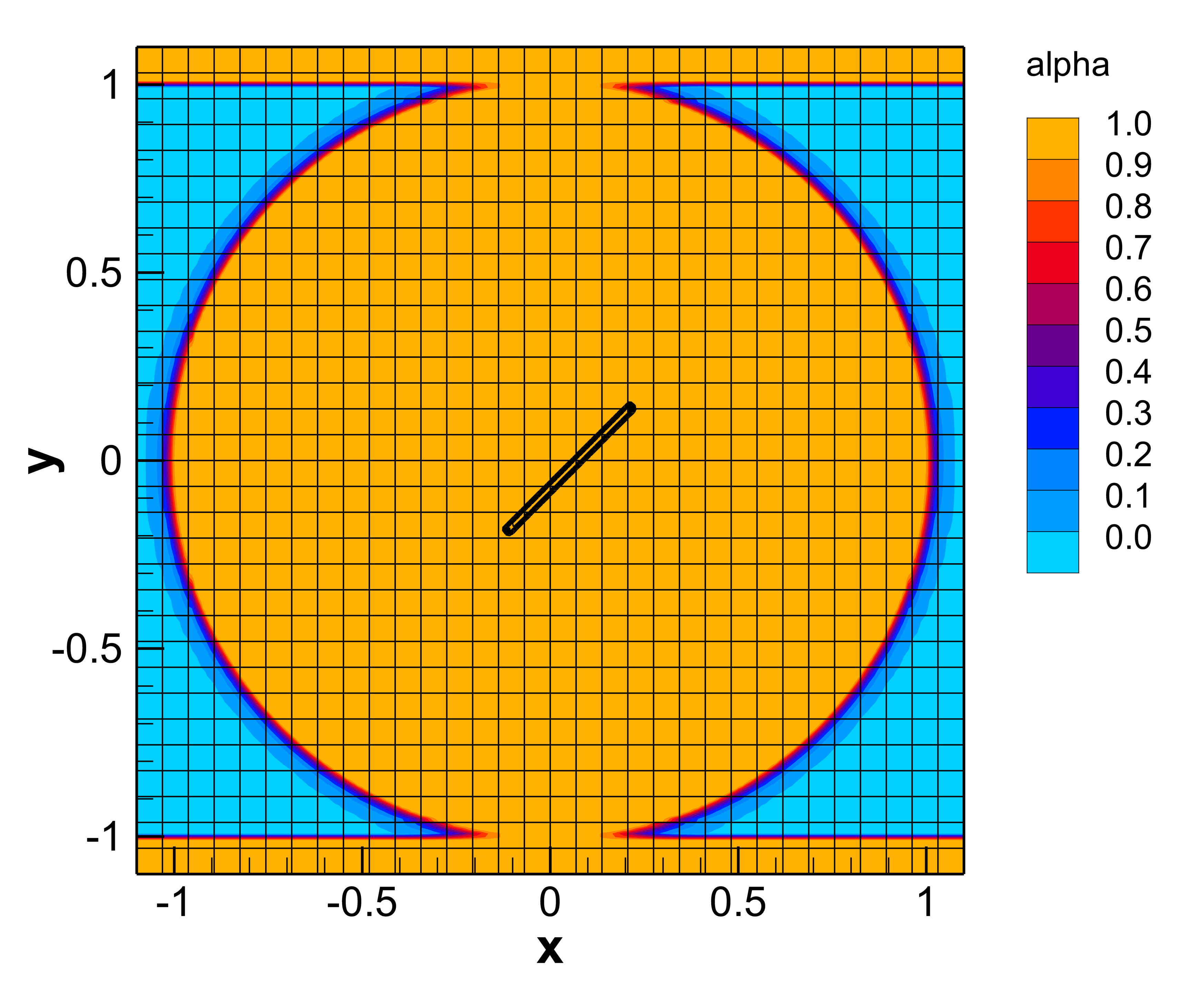}%
		\includegraphics[draft=false,width=0.33\textwidth]{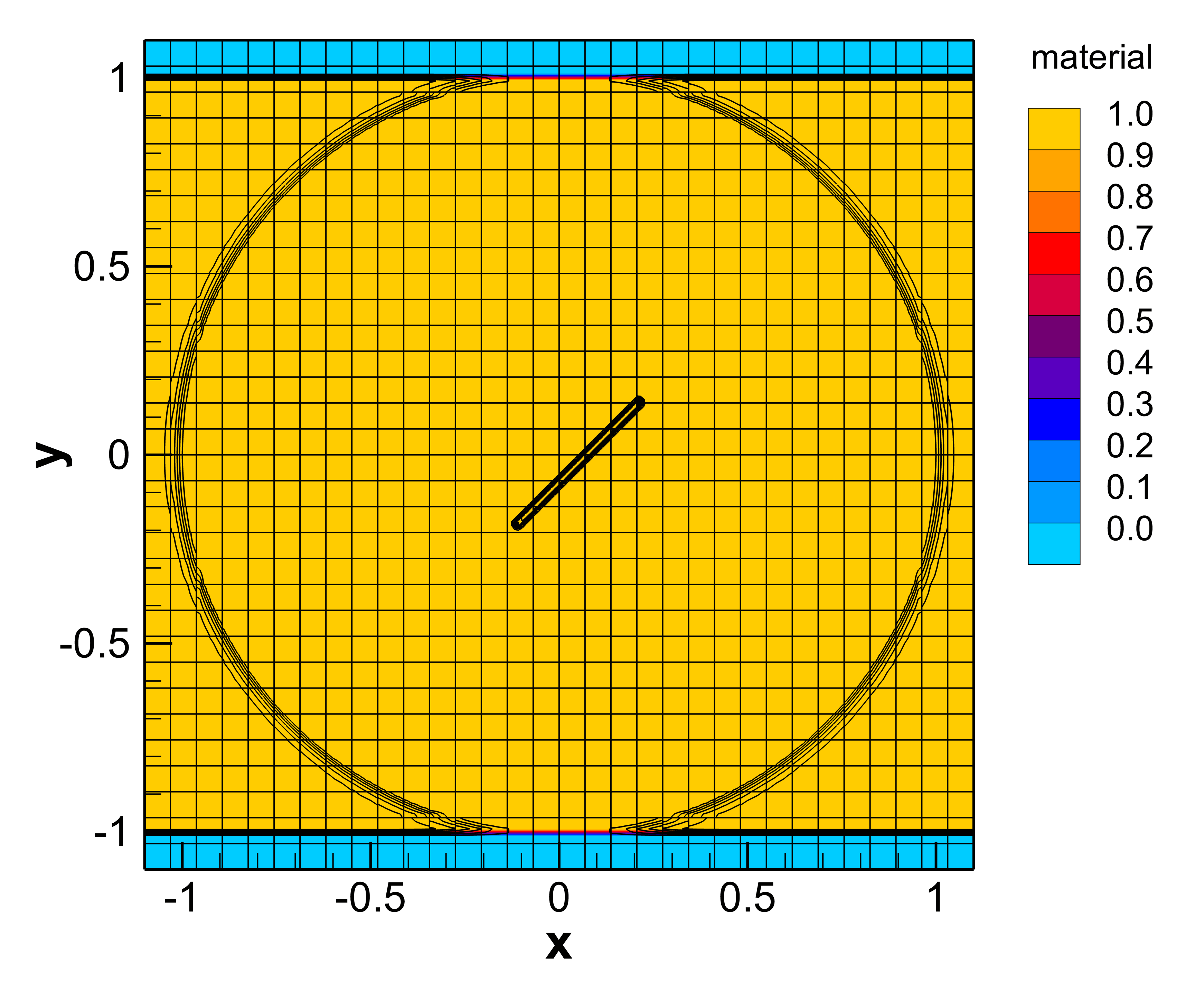}%
	}
\end{minipage}

In \Fig\ref{fig.S5} we observe that a three-fold increase in the mesh resolution 
(from 64 
cells
per space dimension, to 96, to 128, to 192) allows to achieve better sharpness and definition of the
cracks, but leaves their position essentially unaltered. For the last two steps, the crack topology
converges to a stable configuration also at the points of contact between the clamps and the sample,
which appear to be more sensitive to grid resolution with respect to the interior of the disc. 
Furthermore,
in no case we are able to observe any Cartesian mesh-imprinting artefacts in our solutions.
\\

\begin{minipage}[t]{1.0\textwidth}
	\centering
		\includegraphics[draft=false,width=0.9\textwidth]{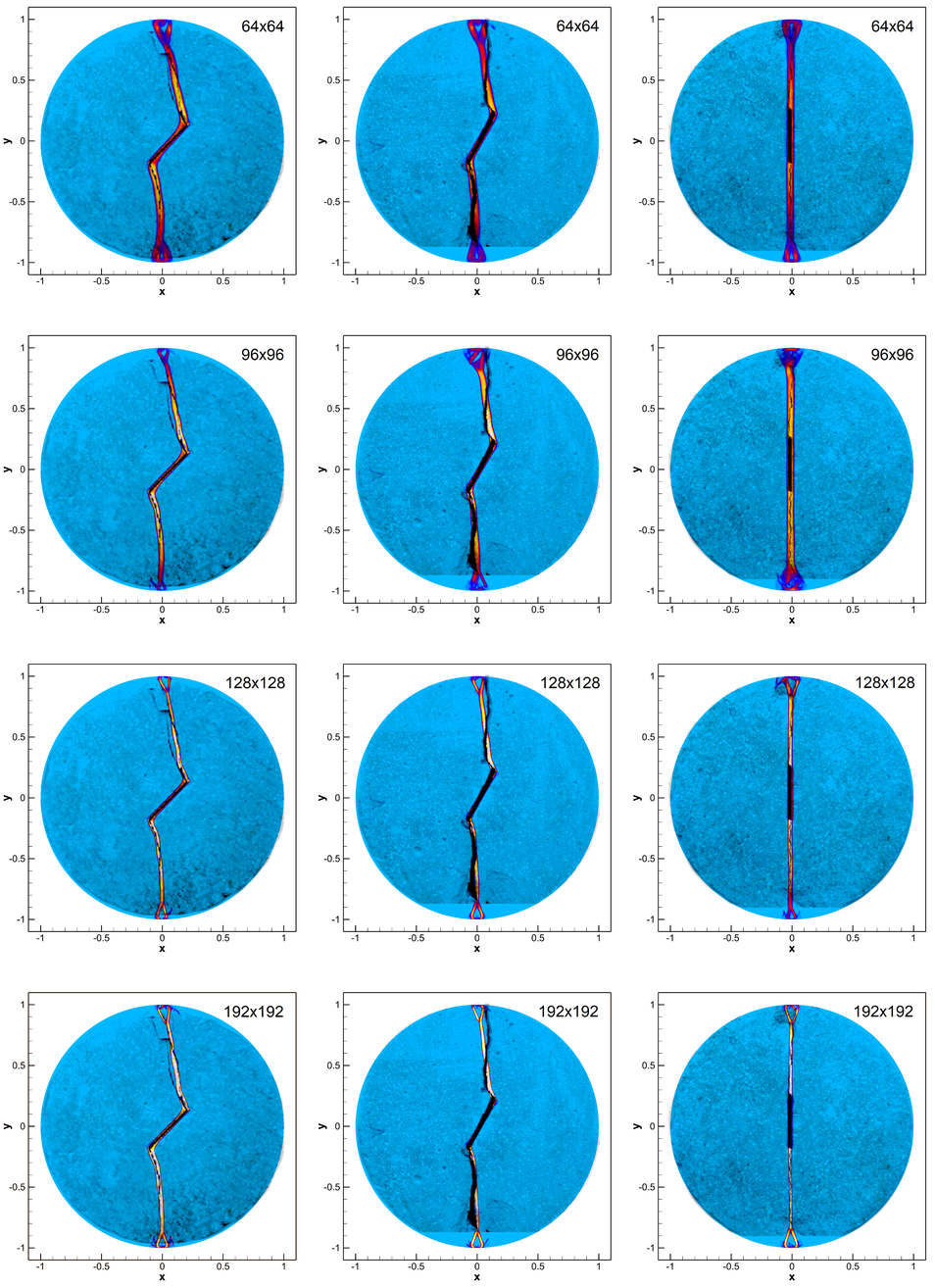}  
	\captionof{figure}{Mesh convergence study for crack formation in a rock-like disc under
		vertical load (Brazilian test). We also compare the contour colours of the damage parameter 
		$ \xi $ obtained
		in the numerical simulations of the GPR model with the cracks observed in experiments.
		From left to right, the inclination angles of the pre-damaged slit are: $ 45^\circ $, $ 
		60^\circ $ and $90^\circ $. 
		From top to
		bottom, the number of grid cells in each dimension is: 64, 96, 128, and 192.} 
	\label{fig.S5}
\end{minipage}

\end{document}

%% file: defs_arxiv.tex
\usepackage[svgnames]{xcolor} 
\usepackage{titling}
\usepackage{fourier}

\newcommand{\Fig}{Fig.}

\usepackage{amsbsy,amssymb,amsmath,setspace,amsfonts}
\usepackage{amsmath}
\usepackage[mathscr]{eucal}
\usepackage{bm}
\usepackage{cite}
\allowdisplaybreaks
\usepackage{comment}
\usepackage{relsize} 
\usepackage[font=small,labelfont=bf]{caption}	

\usepackage{makecell}
\usepackage{floatrow}

\usepackage{graphicx,color}

\usepackage[backref=page]{hyperref} 
\hypersetup{
	colorlinks=true,                          
	linkcolor=Navy, 
	citecolor=DarkRed, 
	urlcolor=blue  }
\renewcommand*{\backref}[1]{} 
\renewcommand*{\backrefalt}[4]{%
    \ifcase #1 (Not cited.)%
    \or        (Cited on page~#2.)%
    \else      (Cited on pages~#2.)%
    \fi} 
\renewcommand{\AA}{{\boldsymbol{A}}}

\newcommand{\vv}{{\boldsymbol{v}}}

\newcommand{\JJ}{{\boldsymbol{J}}}
\newcommand{\II}{{\boldsymbol{I}}}

\newcommand{\pd}{\partial}

\renewcommand{\i}{{ I}}
\renewcommand{\d}{{ D}}

\renewcommand{\i}{{\mathsmaller I}}
\renewcommand{\d}{{\mathsmaller D}}

\newcommand*\samethanks[1][\value{footnote}]{\footnotemark[#1]}

\usepackage[english]{babel}
\usepackage[space, extendedchars]{grffile}
\newcommand{\up}[1]{\ensuremath{\mathrm{#1}}} 

\newcommand\blfootnote[1]{%
	\begingroup
	\renewcommand\thefootnote{}\footnote{#1}%
	\addtocounter{footnote}{-1}%
	\endgroup
}